\documentclass{emulateapj}   
\usepackage{xspace,graphics,eso-pic,color,graphicx,type1cm,ulem,amsmath,color}

\shorttitle{The Y dwarfs}
\shortauthors{Cushing et al.}

\newcommand\teff{\mbox{$T_\mathrm{eff}$}}

\newcommand\logg{\mbox{$\log g$}}

\newcommand\kzz{\mbox{$K_\mathrm{zz}$}}

\newcommand{\ULASzerozerothirtyfourFull}{\mbox{ULAS J003402.77$-$005206.7}}
\newcommand{\ULASzerozerothirtyfour}{\mbox{ULAS J0034$-$0052}}

\newcommand{\ULASthirteenthirtyfiveFull}{\mbox{ULAS J133553.45$+$113005.2}}
\newcommand{\ULASthirteenthirtyfive}{\mbox{ULAS J1335$+$1130}}

\newcommand{\UGPSzeroseventwotwoFull}{\mbox{UGPS J072227.51$-$054031.2}}
\newcommand{\UGPSzeroseventwotwo}{UGPS 0722$-$05}

\newcommand{\CFHBDSzerozerofivenineFull}{\mbox{CFBDS J005910.90$-$011401.3}}
\newcommand{\CFHBDSzerozerofivenine}{\mbox{CFBDS J0059$-$0114}}

\newcommand{\TfourFull}{\mbox{2MASS J22541892$+$3123498}}

\newcommand{\TsixFull}{\mbox{SDSS J162414.37$+$002915.6}}

\newcommand{\TsevenFull}{\mbox{2MASS J07271824$+$1710012}}

\newcommand{\TeightFull}{\mbox{2MASS J04151954$-$0935066}}

\newcommand{\WzeroonefoureightFull}{\mbox{WISEPC J014807.25$-$720258.8}}
\newcommand{\Wzeroonefoureight}{\mbox{WISEPC J0148$-$7202}}

\newcommand{\WzerofourtenFull}{\mbox{WISEP J041022.71$+$150248.5}}
\newcommand{\Wzerofourten}{\mbox{WISEP J0410$+$1502}}

\newcommand{\WfourteenzerofiveFull}{\mbox{WISEPC J140518.40$+$553421.5}}
\newcommand{\Wfourteenzerofive}{\mbox{WISEPC J1405$+$5534}}

\newcommand{\WfifteenfortyoneFull}{\mbox{WISEP J154151.65$-$225025.2}}
\newcommand{\Wfifteenfortyone}{\mbox{WISEP J1541$-$2250}}

\newcommand{\WtwentyfiftysixFull}{\mbox{WISEPC J205628.90$+$145953.3}}
\newcommand{\Wtwentyfiftysix}{\mbox{WISEPC J2056$+$1459}}

\newcommand{\WzerofourfiveeightFull}{\mbox{WISEPC J045853.90$+$643451.9}}

\newcommand{\WseventeenthirtyeightFull}{\mbox{WISEP J173835.52+273258.9}}
\newcommand{\Wseventeenthirtyeight}{\mbox{WISEP J1738+2732}}

\newcommand{\WeighteentwentyeightFull}{\mbox{WISEP J182831.08+265037.8}}
\newcommand{\Weighteentwentyeight}{\mbox{WISEP J1828+2650}}

\newcommand{\CFHfourteenfiftyeightFull}{CFBDSIR J145829$+$101343}
\newcommand{\CFHfourteenfiftyeight}{CFBDSIR J1458$+$1013}

\newcommand{\WDzeroeightzerosixFull}{WD 0806$-$661}

\newcommand{\ULAStwelvethirtyeightFull}{ULAS J123828.51$+$095351.3}
\newcommand{\ULAStwelvethirtyeight}{ULAS J1238$+$0953}

\begin{document}


\title{The Discovery of Y Dwarfs Using Data from the Wide-field Infrared Survey
  Explorer (WISE)}


\author{Michael C. Cushing\altaffilmark{a}, 
  J. Davy Kirkpatrick\altaffilmark{b},
  Christopher R. Gelino\altaffilmark{b},
  Roger L. Griffith\altaffilmark{b},
  Michael F. Skrutskie\altaffilmark{c},
  Amanda K. Mainzer\altaffilmark{d},
  Kenneth A. Marsh\altaffilmark{b},
  Charles A. Beichman\altaffilmark{b},
  Adam J. Burgasser\altaffilmark{e,f},
  Lisa A. Prato\altaffilmark{g},
  Robert A. Simcoe\altaffilmark{f},
  Mark S. Marley\altaffilmark{h},
  D. Saumon\altaffilmark{i},
  Richard S. Freedman\altaffilmark{h},
  Peter R. Eisenhardt\altaffilmark{d}, \& 
  Edward L. Wright\altaffilmark{j}}

%\author{Copyright 2011.  All rights reserved.}

\altaffiltext{a}{Jet Propulsion Laboratory, California Institute of Technology 4800 Oak Grove Drive, MS 321-520, Pasadena, CA 91109; michael.cushing@gmail.com}

\altaffiltext{b}{Infrared Processing and Analysis Center, California
   Institute of Technology, Pasadena, CA 91125}

\altaffiltext{c}{Department of Astronomy, University of Virginia, Charlottesville, VA, 22904}

\altaffiltext{d}{Jet Propulsion Laboratory, California Institute of Technology 4800 Oak Grove Drive, Pasadena, CA 91109}

\altaffiltext{e}{Center for Astrophysics and Space Science, University of California San Diego, La Jolla, CA 92093, Hellman Fellow}

\altaffiltext{f}{Massachusetts Institute of Technology, 77 Massachusetts Avenue, Building 37, Cambridge, MA 02139}

\altaffiltext{g}{Lowell Observatory, 1400 West Mars Hill Road, Flagstaff, AZ, 86001}

\altaffiltext{h}{NASA Ames Research Center, MS 254-3, Moffett Field, CA 94035}

\altaffiltext{i}{Los Alamos National Laboratory, MS F663, Los Alamos, NM
  87545}

\altaffiltext{j}{Department of Physics and Astronomy, UCLA, Los Angeles, CA 90095}

\begin{abstract}

  We present the discovery of seven ultracool brown dwarfs identified
  with the Wide-field Infrared Survey Explorer (WISE).  Near-infrared
  spectroscopy reveals deep absorption bands of H$_2$O and CH$_4$ that
  indicate all seven of the brown dwarfs have spectral types later than
  \UGPSzeroseventwotwoFull, the latest type T dwarf currently known.
  The spectrum of \WeighteentwentyeightFull\ is distinct in that the
  heights of the $J$- and $H$-band peaks are approximately equal in
  units of $f_\lambda$, so we identify it as the archetypal member of
  the Y spectral class.  The spectra of at least two of the other brown
  dwarfs exhibit absorption on the blue wing of the $H$-band peak that
  we tentatively ascribe to NH$_3$.  These spectral morphological
  changes provide a clear transition between the T dwarfs and the Y
  dwarfs. In order to produce a smooth near-infrared spectral sequence
  across the T/Y dwarf transition, we have reclassified
  \UGPSzeroseventwotwo\ as the T9 spectral standard and tentatively
  assign \WseventeenthirtyeightFull\ as the Y0 spectral standard.  In
  total, six of the seven new brown dwarfs are classified as Y dwarfs:
  four are classified as Y0, one is classified as Y0 (pec?), and
  \Weighteentwentyeight\ is classified as $>$Y0.  We have also compared
  the spectra to the model atmospheres of Marley and Saumon and infer
  that the brown dwarfs have effective temperatures ranging from 300 K
  to 500 K, making them the coldest spectroscopically confirmed brown
  dwarfs known to date.

\end{abstract}

\section{Introduction}

Brown dwarfs, objects with too little mass to sustain the high core
temperatures necessary for stable thermonuclear fusion of hydrogen, are
the least massive, and possibly the most numerous products of star
formation.  Although first predicted to exist in the early 1960's
\citep{1963ApJ...137.1121K,1963PThPh..30..460H}, it was not until
decades later that brown dwarfs were discovered in bulk by wide-area,
red (700$-$1000 nm) and near-infrared (1$-$2.5 $\mu$m) capable surveys
such as the Two Micron All Sky Survey
\citep[2MASS,][]{2006AJ....131.1163S}, the Sloan Digital Sky Survey
\citep[SDSS,][]{2000AJ....120.1579Y} and the Deep Near-Infrared Southern
Sky Survey \citep[DENIS,][]{1997Msngr..87...27E}.  The emergent spectra
of brown dwarfs are so distinct from that of late-type M dwarfs that the
creation of two new spectral classes, L and T
\citep{1999ApJ...519..802K,2006ApJ...637.1067B}, was required in order
to properly classify them\footnote{A compendium of known L and T dwarfs
  can be found at \url{http://DwarfArchives.org}}.  The latest type T
dwarfs currently known were discovered in the UKIRT Infrared Deep Sky
Survey \citep[UKIDSS,][]{2007MNRAS.379.1599L} and the Canada France
Brown Dwarf Survey \citep[CFBDS,][]{2008A&A...484..469D} and have
estimated effective temperatures (\teff s) of 500 to 700 K
\citep[e.g.,][]{2008MNRAS.391..320B,2008A&A...482..961D,2009ApJ...695.1517L,2010MNRAS.408L..56L,2011arXiv1103.0014L}.

Despite these spectacular successes, there exists a gap of nearly 400 K
between the coolest spectroscopically confirmed brown dwarfs at \teff\
$\sim$ 500 K \citep{2010MNRAS.408L..56L} and Jupiter at \teff\ $\sim$
124 K \citep{1981JGR....86.8705H}.  Although observations of star
formation regions and young associations such as the Orion Nebula
Cluster \citep{2009MNRAS.392..817W}, Chameleon I
\citep{2005ApJ...635L..93L}, and TW Hydrae \citep{2004A&A...425L..29C}
suggest that nature can form brown dwarfs that will eventually cool to
these temperatures once they have dispersed from their natal clusters,
they are simply too faint to have been detected by the aforementioned
surveys.  Recently however, two brown dwarfs with estimated effective
temperatures of 300$-$400 K, \WDzeroeightzerosixFull B
\citep{2011ApJ...730L...9L} and \CFHfourteenfiftyeightFull B
\citep[\CFHfourteenfiftyeight B][]{2011arXiv1103.0014L}, were discovered
in targeted searches for companions to nearby stars.  Although efforts
to obtain spectra of these two common proper motion sources have been
hampered by extreme faintness in the case of \WDzeroeightzerosixFull
\citep[$J$ $>$ 21.7;][]{2011ApJ...732L..29R}, and proximity to its
primary star in the case of \CFHfourteenfiftyeight B ($<$0$\farcs$1),
their mere existence suggests that a population of nearby cold brown
dwarfs awaits discovery.

Foremost is the question of what these objects will look like
spectroscopically and whether a new spectral class beyond T, dubbed
``Y'' \citep{1999ApJ...519..802K,2000ASPC..212...20K}, will be required
in order to properly classify them.  Chemical equilibrium calculations
and model atmospheres predict that as brown dwarfs cool below \teff\
$\sim$ 600 K, their atmospheres pass through a series of chemical
transitions which in turn impact the appearance of their emergent
spectra
\citep{1999ApJ...519..793L,1999ApJ...512..843B,2002Icar..155..393L,2003ApJ...596..587B}.
At \teff\ $\sim$ 600 K, the combination and overtone bands of NH$_3$
emerge in the near-infrared\footnote{Although the fundamental band of
  NH$_3$ has been detected in the spectra of warmer T dwarfs at 10.5
  $\mu$m \citep[e.g.,][]{2004ApJS..154..418R,2006ApJ...648..614C}, the
  intrinsically weaker near-infrared bands require a higher NH$_3$
  abundance, and thus lower effective temperature, to become a dominant
  opacity source.}.  At \teff\ $\sim$ 500 K, the prominent resonance
absorption lines of \ion{Na}{1} and \ion{K}{1} in the red optical
spectra of warmer brown dwarfs weaken as Na condenses out of the gas
phase into Na$_2$S and then K condenses into KCl.  Finally, H$_2$O and
NH$_3$ will also condense out at \teff\ $\sim$ 350 K and $\sim$200 K,
respectively.  Although each transition has been suggested as the
trigger for the Y spectral class, focus has primarily been on detecting
the NH$_3$ bands because they are predicted to emerge at the hottest
effective temperatures.  Although NH$_3$ absorption has been tentatively
detected in the near-infrared spectrum of \CFHBDSzerozerofivenineFull\
\citep[hereafter \CFHBDSzerozerofivenine,][]{2008A&A...482..961D}, this
feature has not been confirmed in the spectrum of the cooler object
\UGPSzeroseventwotwoFull\
\citep[\UGPSzeroseventwotwo;][]{2010MNRAS.408L..56L}.

Independent of their spectral morphology, the study of these ultracool
brown dwarfs will provide important insights into both ultracool
atmospheric physics and the low-mass end of the stellar mass function.
Because brown dwarfs and exoplanets have similar atmospheric conditions,
ultracool brown dwarfs are also excellent exoplanet analogs that can be
used as benchmarks for model atmospheres.  The study of these ultracool
brown dwarfs will therefore directly inform the interpretation and
characterization of exoplanets detected with the next generation of
high-contrast imagers like the Gemini Planet Imager
\citep[GPI;][]{2006SPIE.6272E..18M} the Spectro-Polarimetric
High-contrast Exoplanet Research instrument for the VLT
\citep[SPHERE;][]{2006tafp.conf..353B}, Project 1640 at Palomar
Observatory \citep{2011PASP..123...74H}, and the $L$- and $M$-band
Infrared Camera \cite[LMIRcam;][]{2010SPIE.7735E.118S} for the Large
Binocular Telescope Interferometer (LBTI).  Simulations by
\citet{2004ApJS..155..191B} and \citet{2005ApJ...625..385A} have also
shown that the space density of cold brown dwarfs is very sensitive to
both the underlying mass function and the low-mass limit of star
formation.  Identifying and characterizing a statistically robust sample
of cold brown dwarfs will therefore provide two critical constraints on
theories of low-mass star formation
\citep{2006A&A...458..817W,2007prpl.conf..459W}.

One of the primary science goals of the Wide-field Infrared Survey
Explorer \citep[WISE,][]{2010AJ....140.1868W}, a NASA mission that
recently surveyed the entire sky at 3.4 (W1), 4.6 (W2), 12 (W3), and 22
(W4) $\mu$m, is to identify such cold brown dwarfs.  The W1 and W2 bands
were designed specifically to sample the deep CH$_4$ absorption band
centered at 3.3 $\mu$m and the region relatively free of opacity
centered at $\sim$4.7 $\mu$m in the spectra of cold brown dwarfs
\citep[see Figure 2 of][]{2011ApJ...726...30M}.  Since the peak of the
Planck function at these low effective temperatures is in the
mid-infrared, a large amount of flux emerges from the 4.7 $\mu$m opacity
hole, making the W1$-$W2 colors extremely red \citep[W1$-$W2 $>$
2,][]{2011ApJ...726...30M,Davy11}. Indeed such red colors are almost
unique amongst astronomical sources making the identification of cool
brown dwarfs with the W1$-$W2 color alone relatively easy \citep[see
Figure 12 of][]{2010AJ....140.1868W}.

We have been conducting a search for cold brown dwarfs since the start
of the WISE survey in mid January 2010.  This search has already born
fruit with the discovery of six late-type T dwarfs
\citep{2011ApJ...726...30M,2011ApJ...735..116B} two of which have
spectral types later than T8.  \citet{Davy11} present over 100 new brown
dwarfs, the vast majority of which have spectral types later than T6. In
this paper, we focus on seven of the $\sim$100 brown dwarfs whose
near-infrared spectra indicate that the they are the latest-type
spectroscopically confirmed brown dwarfs currently known.  Indeed we
identify six of these brown dwarfs as the first members of the Y
spectral class.  In \S2, we briefly discuss our selection criteria
before presenting the ground- and spaced-based imaging and spectroscopic
followup observations in \S3.  In \S4 we present the properties of the
first Y dwarfs, define the transition between the T sequence and the Y
dwarfs, and derive estimates of the atmospheric parameters of the new
brown dwarfs using model atmospheres.

\section{Candidate Selection}

The seven new brown dwarfs were identified as part of a larger ongoing
search for cold brown dwarfs using WISE.  A detailed description of this
survey and our search criteria is presented by \citet{Davy11}.  Briefly,
candidates were selected from the source working database derived from
the first-pass processing operational coadds using color constraints
derived from known T dwarfs and model atmospheres (in the case of brown
dwarfs with \teff\ $<$ 500 K).  Table \ref{tab:wisephot} lists the WISE
designations and photometry of the seven brown dwarfs, and Figure 1
shows 2$'$ x 2$'$ DSS I, 2MASS $J$ and $H$, WISE W1, W2, and W3, and
W1W2W3 color composite images for each dwarf.  Hereafter we abbreviate
the numerical portions of the WISE designations as hhmm$\pm$ddmm, where
the suffix is the sexagesimal right ascension (hours and minutes) and
declination (degrees and arcminutes) at J2000.0 equinox.

\section{Observations}

The followup ground- and space-based observations of the seven WISE
brown dwarfs are discussed in the following sections.  Although we
present the near-infrared photometry of the brown dwarfs in this work
for completeness, we defer the discussion of these data to
\citet{Davy11} who present a more detailed discussion that places them
in context with the larger population of brown dwarfs. In addition to
the observations of the seven WISE brown dwarfs, we also obtained a
near-infrared spectrum of \UGPSzeroseventwotwo\ for comparison
purposes. A log of the near-infrared photometric observations as well as
the resulting photometry is given in Table \ref{tab:nirphot}, and a log
of the spectroscopic observations is given in Table \ref{tab:speclog}.

\subsection{Near-Infrared Imaging}

\subsubsection{NEWFIRM/Blanco}

\Wfifteenfortyone\ was observed on the night of 2011 Apr 17 (UT) with
the NOAO Extremely Wide Field Infrared Imager (NEWFIRM) mounted on the
Cerro Tololo Inter-American Observatory (CTIO) Victor M. Blanco 4m
Telescope.  A description of the instrument, observing strategy, and
data reduction can be found in \citet{Davy11}.  The resultant $J$- and
$H$-band photometry is presented in Table \ref{tab:nirphot}.

\subsubsection{WIRC/Palomar}
\label{sec:wirc}

Near-infrared images of \Wzerofourten, \Wfourteenzerofive,
\Wseventeenthirtyeight, and \Wtwentyfiftysix\ were obtained using the
Wide-field Infrared Camera \citep[WIRC; ][]{2003SPIE.4841..451W} on the
200 inch Hale Telescope at Palomar Observatory.  A description of the
instrument, observing strategy, and data reduction can be found in
\citet{Davy11}.  The magnitudes and/or limits for each brown dwarf are
given in Table \ref{tab:nirphot}.

\subsubsection{PANIC/Magellan}

\Wzeroonefoureight\ was observed on the night of 2010 Aug 01 (UT) with
the now decommissioned Persson's Auxiliary Nasmyth Infrared Camera
\citep[PANIC; ][]{2004SPIE.5492.1653M} on the east Nasmyth platform at
the Magellan 6.5 m Baade Telescope.  A description of the instrument,
observing strategy, and data reduction can be found in \citet{Davy11}.
The $J$- and $H$-band magnitudes of \Wzeroonefoureight\ are given in
Table \ref{tab:nirphot}.

\subsubsection{NIRC2/Keck II}

High resolution observations of \Weighteentwentyeight\ and
\Wtwentyfiftysix\ were obtained with NIRC2 behind the Keck II LGS-AO
system \citep{2006PASP..118..297W,2006PASP..118..310V} on the night of
2010 July 1 (UT).  A description of the instrument, observing strategy,
and data reduction can be found in \citet{Davy11}.  The $J$- and
$H$-band magnitudes are given in Table \ref{tab:nirphot}.

\subsection{Near-Infrared Spectroscopy}

\subsubsection{SpeX/IRTF}
\label{sec:SpeX}

A 0.9$-$2.5 $\mu$m, low-resolution ($R\equiv$ $\lambda / \Delta \lambda$
$\approx$ 150) spectrum of \UGPSzeroseventwotwo\ was obtained with SpeX
\citep{2003PASP..115..362R} on the 3 m NASA IRTF on 26 Jan 2011 (UT).  A
description of the instrument, observing strategy, and data reduction
can be found in \citet{Davy11}.  The spectrum, which is shown in Figure
\ref{fig:rawseq}, has a high S/N, reaching $>$50 at the peaks of the
$Y$, $J$, and $H$ bands.

\subsubsection{FIRE/Magellan}

Low-resolution ($R$=250$-$350), 1$-$2.4 $\mu$m spectra of
\Wzeroonefoureight, \Wzerofourten, and \Wfifteenfortyone\ were obtained
with the Folded-port InfraRed Echellette
\citep[FIRE,][]{2008SPIE.7014E..27S,2010SPIE.7735E..38S} mounted at the
auxiliary Nasmyth focus of the Magellan 6.5 m Baade Telescope . A
description of the instrument, observing strategy, and data reduction
can be found in \citet{Davy11}.  The spectra are shown in Figure
\ref{fig:rawseq}.

\subsubsection{NIRSPEC/Keck II}

\Wtwentyfiftysix\ was observed using the Near-Infrared Spectrometer
\citep[NIRSPEC;][]{1998SPIE.3354..566M,2000SPIE.4008.1048M} located on
one of the Nasmyth platforms of the 10 m Keck II telescope on Mauna Kea,
Hawai'i.  The 0$\farcs$38-wide slit in the low resolution mode provides
a resolving power of $R$=2500. \Wtwentyfiftysix\ was observed with the
N3 order sorting filter (1.143$-$1.375 $\mu$m) on the night of 2010 Oct
21 (UT) and with the N5 order sorting filter (1.431$-$1.808 $\mu$m) on
the night of 2010 Nov 22 (UT).

A series of 300 s exposures was obtained at two different positions
along the 42$\arcsec$ long slit.  An A0 V star was observed after each
series of science exposures for telluric correction and flux calibration
purposes.  Calibration frames consisting of neon and argon arc lamps,
dark frames, and flat-field lamps were also taken following the science
exposures.  The data were reduced in a standard fashion using the
IDL-based REDSPEC\footnote{See
  \url{http://www2.keck.hawaii.edu/inst/nirspec/redspec}.}  reduction
package as described in \citet{2003ApJ...596..561M}.  Since REDSPEC does
not produce uncertainty arrays, we generated them as follows.  First we
performed a simple sum extraction using the rectified, pair-subtracted
images generated by REDSPEC.  We then scaled the spectra to a common
flux level and computed the average spectrum.  The uncertainty at each
wavelength is given by the standard error on the mean.  The average
spectrum is then corrected for telluric absorption and flux calibrated
using the calibration spectrum generated by REDSPEC.  Since the
difference between the spectra produced by REDSPEC and our spectra was
negligible, we used our spectrum for our analysis.  Finally, the N3$-$
and N5$-$band spectra were absolutely flux calibrated using the WIRC
photometry (see Table \ref{tab:nirphot}) as described in
\citet{2005ApJ...623.1115C} and merged to produce a 1.15$-$1.80 $\mu$m
spectrum. The final spectrum, which is shown in Figure \ref{fig:rawseq},
has a peak S/N of 8 and 6 in the $J$ and $H$ bands, respectively.

\subsubsection{WFC3/Hubble Space Telescope}
\label{sec:wfc3}

\Wfourteenzerofive, \Wseventeenthirtyeight, and \Weighteentwentyeight\
were observed with the infrared channel of the Wide Field Camera 3
\citep[WFC3;][]{2008SPIE.7010E..43K} on-board the \textit{Hubble Space
  Telescope} (HST) as a part of a Cycle 18 program (GO-12330,
PI=Kirkpatrick).  The WFC3 uses a 1024 $\times$ 1024 HgCdTe detector
with a plate scale of 0$\farcs$13 pixel$^{-1}$ which results in a field
of view of 123$\times$126 arcsecond.  The G141 grism was used to perform
slitless spectroscopy of each brown dwarf covering the 1.07$-$1.70
$\mu$m wavelength range at a resolving power of $R$$\approx$130.  For
each brown dwarf, we first obtained four direct images through the F140W
filter ($\lambda_p$=1392.3 nm) in the MULTIACCUM mode with the SPARS25
sampling sequence.  Between each exposure the telescope was offset
slightly.  We then obtained four images with the G141 grism at the same
positions as the direct images. The spectroscopic observations were also
obtained in the MUTLIACCUM mode but using the SPARS50 sequence.

The raw images were first processed using the CALWFC3 pipeline (version
2.3) which not only subtracts the bias level and dark current but also
flat fields the direct images (the grism images are flat fielded during
the extraction process described below).  The spectra were then
extracted using the aXe software \citep{2009PASP..121...59K}, which is a
suite of PyRAF/IRAF packages designed to extract spectra from the
slitless modes of both WFC3 and the Advanced Camera for Surveys (ACS).
aXe requires knowledge of both the position and brightness of the
objects in the field of view.  We therefore combined the four direct
images using MULTIDRIZZLE \citep{2002hstc.conf..337K} and the latest
Instrument Distortion Coefficient Table (IDCTAB).  A catalog of the
objects in the field was then constructed using SExtractor
\citep{1996A&AS..117..393B}.  For each object in the source catalog, two
dimensional (2D) subimages centered on the first-order spectra of each
object were then combined using the task AXEDRIZZLE to produce a high
S/N 2D spectral image.  One-dimensional, flux-calibrated spectra and
their associated uncertainties are then extracted from the 2D drizzle
subimages.

Since the G141 grism mode is slitless, spectral contamination from
nearby sources is not uncommon.  The aXe software (using the Gaussian
emission model) estimates the level of contamination for each object
using the positions and magnitudes of all the objects in the field of
view.  The spectrum of one of the brown dwarfs, \Weighteentwentyeight,
exhibits moderate contamination that increases in intensity towards
shorter wavelengths (see Figure \ref{fig:1828img}).  The aXe software
does not actually correct for this contamination so we attempted to do
so using the contamination image generated by aXe.  Unfortunately, the
contamination-corrected spectrum exhibits negative flux values which
suggests that aXe is over estimating the contamination level. We will
therefore use the contaminated spectrum and consider it an upper limit
to the actual spectrum.  This issue will be discussed in more detail in
\S\ref{sec:1828}.

\section{Analysis}

\subsection{Spectral Characterization}

Figure \ref{fig:rawseq} shows the near-infrared spectra of the new brown
dwarfs.  Also plotted for comparison purposes is our IRTF/SpeX spectrum
of \UGPSzeroseventwotwo, the latest type brown dwarf known previous to
this work.  All of the spectra exhibit deep H$_2$O and CH$_4$ absorption
bands characteristic of late-type T dwarfs but the $J$-band peaks of the
WISE brown dwarfs are narrower than the corresponding peak in the
spectrum of \UGPSzeroseventwotwo.  This peak becomes progressively
narrower beyond T8
\citep{2007MNRAS.381.1400W,2008A&A...482..961D,2008MNRAS.391..320B,2010MNRAS.408L..56L},
indicating that all of the WISE brown dwarfs have spectral types later
than \UGPSzeroseventwotwo.  The spectrum of \Weighteentwentyeight\ is
markedly different than that of \UGPSzeroseventwotwo\ so we discuss this
object in more detail in the following section before discussing the
other six dwarfs.

\subsubsection{\Weighteentwentyeight: The Archetypal Y Dwarf}

\label{sec:1828}

The lower panel of Figure \ref{fig:t9y0comp} shows the 1.15$-$1.70
$\mu$m spectrum of \Weighteentwentyeight\ along with the spectrum of
\UGPSzeroseventwotwo.  The spectrum of \Weighteentwentyeight, while
dominated by the same CH$_4$ and H$_2$O absorption bands present in T
dwarf spectra, has a feature not seen in any T dwarf: the $J$- and $H$-
band peaks, when plotted in units of $f_\lambda$, are essentially the
same height. As discussed in \S\ref{sec:wfc3}, the spectrum of
\Weighteentwentyeight\ is contaminated by light from nearby stars.  This
contamination, which is \textit{not} removed by the aXe software,
becomes progressively worse at shorter wavelengths (see Figure
\ref{fig:1828img}) such that the true spectrum will have an even more
extreme $J$ to $H$ band peak flux ratio.

The roughly equal-intensity $J$ and $H$ flux peaks are also confirmed by
our ground-based, near-infrared photometry, which gives $J-H =
0.72{\pm}0.42$ mag.  Model atmospheres of cool brown dwarfs predict that
the near-infrared colors, which are blue for the hotter T dwarfs, turn
back to the red at effective temperatures between 300 and 400 K as the
Wien tail of the spectral energy distribution collapses. This turn to
the red was proposed as one of the triggers that might force the
creation of a Y spectral class
\citep{2003ApJ...596..587B,2008ASPC..384...85K}.

Further underscoring the extreme nature of \Weighteentwentyeight\ is its
$J-$W2 color of 9.29$\pm$0.35 which is over 2 mags redder than the
\Wfifteenfortyone, the second reddest brown dwarf in our sample at
$J-$W2=7.18$\pm$0.38 \citep{Davy11}.  \Weighteentwentyeight\ is also the
reddest brown dwarf in our sample in $H-$W2, $J-$[4.5], and $H-$[4.5],
where [4.5] represents the \textit{Spitzer Space Telescope}
\citep{2004ApJS..154....1W} Infrared Array Camera
\citep[IRAC;][]{2004ApJS..154...10F} channel 2 magnitude.  Given the
extreme nature of both its near-infrared spectrum and near- to
mid-infrared colors, we identify \Weighteentwentyeight\ as the
archetypal member of the Y spectral class.

\subsubsection{\Wseventeenthirtyeight\ \& \Wfourteenzerofive\ and the
  1.5 $\mu$m NH$_3$ Band}

The 1.15$-$1.70 $\mu$m spectra of \Wfourteenzerofive\ and
\Wseventeenthirtyeight\ are very similar and yet both are distinct from
\UGPSzeroseventwotwo, albeit in less extreme ways than the spectrum of
\Weighteentwentyeight\ (see middle and upper panels of Figure
\ref{fig:t9y0comp}).  Although the relative heights of the $J$- and
$H$-band peaks are similar to those of \UGPSzeroseventwotwo, their
widths are narrower.  The narrowing of the $H$-band peaks is asymmetric,
however, as most of the change is a result of enhanced absorption on the
blue wings from 1.51$-$1.58 $\mu$m.  What is the underlying cause of
this absorption?

The $H$-band spectra of T dwarfs are shaped by CH$_4$ (and to a lesser
extent H$_2$O) longward of 1.6 $\mu$m and by H$_2$O at wavelengths
shortward of 1.6 $\mu$m.  As the effective temperature falls, the
opacity of the near infrared overtone and combination bands of NH$_3$
becomes important since NH$_3$/N$_2$ $>$ 1 for $T$ $\lesssim$ 700 K at
$P$=1 bar \citep{2002Icar..155..393L}.  The emergence of these NH$_3$
bands has long been suggested as the trigger for a new spectral class
\citep{2003ApJ...596..587B,2008ASPC..384...85K,2007ApJ...667..537L} but
identifying them has proven difficult because they overlap with the
strong H$_2$O bands and because the abundance of NH$_3$ can be reduced
by an order of magnitude due to vertical mixing in the atmospheres of
brown dwarfs
\citep{2003IAUS..211..345S,2006ApJ...647..552S,2007ApJ...669.1248H}.

Figure \ref{fig:nh3comp} shows the $H$-band spectra of the T4, T6, and
T8 spectral standards, \UGPSzeroseventwotwo, and \Wseventeenthirtyeight\
as well as the opacities for H$_2$O \citep{2008ApJS..174..504F}, NH$_3$
\citep{2011MNRAS.413.1828Y}, and CH$_4$ \citep{2008ApJS..174..504F} at
$T$=600 K and $P$=1 bar generated by one of us by (R.S.F.).  With
increasing spectral type, the blue-wing of the $H$-band peak becomes
progressively suppressed.  \citet{2008A&A...482..961D} tentatively
identified NH$_3$ absorption on the blue wing of the $H$-band spectrum
of \CFHBDSzerozerofivenine, a T dwarf with a spectral type earlier than
\UGPSzeroseventwotwo.  However, the change in the shape of the blue wing
of the $H$-band peak from T6 to \UGPSzeroseventwotwo\ appears smooth,
suggesting a common absorber or set of absorbers.  It seems unlikely
that NH$_3$ dominates given that it has not been identified in the
spectra of mid-type T dwarf (\teff\ $\sim$ 1200 K).  A similar
conclusion to ours is reached by \citet{2010MNRAS.406.1885B} using
spectral indices.  

In contrast, the $H$-band spectrum of \Wseventeenthirtyeight\ stands out
in the sequence in that it exhibits additional absorption from 1.53 to
1.58 $\mu$m.  This absorption broadly matches the position of the
$\nu_1$ $+$ $\nu_3$ absorption band of NH$_3$ centered at 1.49 $\mu$m
suggesting that NH$_3$ is the cause of this absorption. However, we
cannot conclusively identify NH$_3$ as the carrier given the low
spectral resolution of the data and the fact that the absorption lies on
the steep wing of the H$_2$O band.  For example, water ice also has an
absorption band centered at $\sim$1.5 $\mu$m \citep{2008JGRD..11314220W}
that could potentially produce such absorption if the abundance of water
ice is high enough.  One potential avenue for confirming that NH$_3$ is
indeed the carrier would be to acquire higher spectral resolution data
to search for individual NH$_3$ features
\citep[e.g.,][]{2000ApJ...541..374S,2007MNRAS.381.1400W}.

\subsubsection{The T Dwarf/Y Dwarf Transition}
\label{sec:tytransition}

With \Weighteentwentyeight\ classified as the prototypical Y dwarf, we
can now investigate the transition between the T and Y spectral classes.
T dwarfs are classified at near-infrared wavelengths using the
\citet{2006ApJ...637.1067B} scheme, wherein nine T dwarf spectral
standards with subtypes ranging from T0 to T8 are used for direct
spectral comparisons.  \citeauthor{2006ApJ...637.1067B} also defined
five spectral indices that measure the depths of the CH$_4$ and H$_2$O
bands which can be used as a proxy for direct comparisons.  With the
discovery of brown dwarfs with spectral types later than T8, the
question of how to extend the \citeauthor{2006ApJ...637.1067B} scheme
beyond T8 naturally arises.

The first $>$T8 dwarf to be identified was \ULASzerozerothirtyfourFull\
\citep[\ULASzerozerothirtyfour; ][]{2007MNRAS.381.1400W}.  Based on a
direct comparison to the spectrum of the T8 spectral standard and the
values of the \citeauthor{2006ApJ...637.1067B} spectral indices,
\citeauthor{2007MNRAS.381.1400W} adopted a spectral type of T8.5.  A
second $>$T8 dwarf soon followed with the discovery of
\CFHBDSzerozerofivenine\ by \citet{2008A&A...482..961D}.  They used the
W$_J$ index which measures the half-width of the $J$ band peak
\citep{2007MNRAS.381.1400W} and the NH$_3$-$H$ index which measures the
half-width of the $H$-band peak (i.e. the depth of the putative NH$_3$
absorption), to classify both \CFHBDSzerozerofivenine\ and
\ULASzerozerothirtyfour\ as T9 dwarfs.  \citet{2008MNRAS.391..320B}
added two T dwarfs to the tally of $>$T8 dwarfs with the discovery of
\ULASthirteenthirtyfiveFull\ (\ULASthirteenthirtyfive) and
\ULAStwelvethirtyeightFull\ (\ULAStwelvethirtyeight).  Using both direct
spectral comparison and spectral indices, they classified them as T8.5
and T9, respectively.  \citeauthor{2008MNRAS.391..320B} also proposed
extending the \citeauthor{2006ApJ...637.1067B} scheme to T9 by assigning
\ULASthirteenthirtyfive\ as the T9 spectral standard.  Additional T
dwarfs with spectral types later than T8 have since been discovered (see
Table \ref{tab:gtt8}), but to date, the latest-type T dwarf currently
known is \UGPSzeroseventwotwo\ \citep{2010MNRAS.408L..56L} which has
been classified as T10 via a combination of spectral indices and direct
comparisons to the T9 dwarfs.

Figure \ref{fig:t6t7t8t9y0} shows the 1.15$-$1.70 $\mu$m spectra of the
T6, T7, T8 spectral standards, \UGPSzeroseventwotwo, and
\Wseventeenthirtyeight.  The spectra show smooth changes in their
spectral morphology with increasing spectral type including
progressively deeper absorption bands centered at 1.15, 1.45, and 1.65
$\mu$m and progressively narrower $J$- and $H$-band peaks. However
\UGPSzeroseventwotwo\ does not appear to be two subtypes later than T8
as required by its T10 spectral type.  Rather, the changes in spectral
morphology from T6 to \UGPSzeroseventwotwo\ suggest that
\UGPSzeroseventwotwo\ is more naturally classified as a T9.  Given its
brightness ($J$=16.5, 1.5 mag brighter than \ULASthirteenthirtyfive) and
position near the celestial equator, \UGPSzeroseventwotwo\ also makes an
ideal spectral standard.  We therefore define it to be the T9 spectral
standard.

\Wseventeenthirtyeight\ is clearly of later type than
\UGPSzeroseventwotwo\ but should it be classified as a T dwarf or a Y
dwarf?  As noted in the previous section, \Wseventeenthirtyeight\
exhibits excess absorption from 1.53 to 1.58 $\mu$m that we have
tentatively ascribed to NH$_3$.  This absorption becomes even more
apparent when the spectrum is placed in sequence with the T6 to T9
spectral standards (lower right panel of Figure \ref{fig:t6t7t8t9y0}).
Given the smooth change in width of the $J$-band peak and the rapid fall
in the flux of the blue wing of the $H$-band between
\UGPSzeroseventwotwo\ and \Wseventeenthirtyeight\ (which suggests the
emergence of a new absorption band), we classify \Wseventeenthirtyeight\
as a Y dwarf and assign it a spectral type of Y0.  Additionally, we
tentatively identify it as Y0 spectral standard.  

\subsubsection{Classification of the Other WISE Discoveries}

\label{sec:theothers}

With the T9 and Y0 spectral standards defined, we can return to the
question of classifying the other new WISE discoveries.  The $J$- and
$H$-band peaks of \Wzeroonefoureight\ are slightly narrower than
\UGPSzeroseventwotwo\ and slightly wider than \Wseventeenthirtyeight\ so
we classify this dwarf as T9.5.  The spectrum of \Wfourteenzerofive\ is
very similar to that of \Wseventeenthirtyeight\ (see Figure
\ref{fig:t9y0comp}).  However, we note that the wavelength at which the
peak $H$-band flux is reached is shifted $\sim$60 \AA\ to the red
relative to \UGPSzeroseventwotwo\ (see Figure \ref{fig:t9y0comp}) and
the other late-type T dwarfs which suggests that \Wfourteenzerofive\ may
be peculiar.  We therefore classify it as Y0 (pec?).  Interestingly, a
similar, albeit larger, shift of 200 \AA\ is seen in the spectrum of
Jupiter.

The spectra of the remaining brown dwarfs, \Wtwentyfiftysix,
\Wzerofourten, and \Wfifteenfortyone, do not have sufficient S/N to
convincingly show the excess absorption from 1.53$-$1.58 $\mu$m.
However, the $J$-band peaks of these three brown dwarfs are all narrower
than \UGPSzeroseventwotwo.  Indeed the spectra of all of them are a
better match to the spectral morphology of \Wseventeenthirtyeight\ than
\UGPSzeroseventwotwo\ so we classify these brown dwarfs as Y0 as well.
In addition, the peak $Y$-band fluxes of \Wzerofourten\ and
\Wfifteenfortyone\ are slightly higher than in the spectrum of
\UGPSzeroseventwotwo.  This is consistent with the blueward trend in the
$Y-J$ color of late-type T dwarfs
\citep{2010ApJ...710.1627L,2010MNRAS.406.1885B} which
\citet{2010MNRAS.406.1885B} ascribed to the weakening of the strong
resonance \ion{K}{1} doublet (7665, 7699 \AA) as K condenses into KCl.
Finally, since the spectrum of \Weighteentwentyeight\ is distinct from
both \UGPSzeroseventwotwo\ and \Wseventeenthirtyeight, we classify it as
$>$Y0.  A more precise subtype will require the discovery of additional
Y dwarfs to bridge the gap in spectral morphology between
\Wseventeenthirtyeight\ and \Weighteentwentyeight.

\subsubsection{Reclassification  of Previously Published $\ge$T8.5 Dwarfs}

There are also twelve T dwarfs with spectral types later than T8
currently in the literature (see Table \ref{tab:gtt8}).  Since we have
reclassified \UGPSzeroseventwotwo\ as a T9 dwarf, we must also
reclassify the other eleven dwarfs using this new system.  To accomplish
this, we have smoothed the published spectra to a resolving power of
$R$=150 and resampled them onto the same wavelength scale as the
IRTF/SpeX spectrum of \UGPSzeroseventwotwo.  This ensures that
differences in resolving power and wavelength sampling between the
late-type T dwarfs and the \citeauthor{2006ApJ...637.1067B} IRTF/SpeX
spectra of the T dwarf spectral standards do not adversely affect our
classification.  Table \ref{tab:gtt8} gives the revised spectral types
derived from direct comparison for the twelve T dwarfs with published
spectral types later than T8.  When the $J$ and $H$ band regions gave
conflicting spectral types, we used the typed inferred from the
$J$-band.

\subsubsection{Spectral Indices}

Although the primary (and preferred) method of assigning a spectral type
is to compare the spectrum of a brown dwarf against that of the spectral
standards, the use of spectral indices remains popular in the
literature.  We have therefore computed the H$_2$O-$J$, CH$_4$-$J$,
H$_2$O-$H$, CH$_4$-$H$ \citep{2006ApJ...637.1067B}, W$_J$
\citep{2007MNRAS.381.1400W}, and NH$_3$-$H$ \citep{2008A&A...482..961D}
indices of the new WISE brown dwarfs and as well as
\UGPSzeroseventwotwo.  Figure \ref{fig:idxkey} illustrates the positions
of the indices' flux windows relative to the spectrum of
\UGPSzeroseventwotwo.  The index values are computed in a Monte Carlo
fashion whereby 5000 realizations of each spectrum are generated by
randomly drawing from normal distributions with means given by the flux
densities at each wavelength and standard deviations given by the
uncertainty in the flux densities.  The values of the indices and their
uncertainties are given by the mean and standard deviation of the
distribution of index values computed from the 5000 realizations and are
listed in Table \ref{tab:spectralindices}.

Figure \ref{fig:indices} shows the values of the six indices as a
function of spectral type.  Also shown are the index values of the
T6$-$T8 spectral standards, a sample of T5$-$T8 dwarfs from the SpeX
Prism Spectral Library, and the twelve T dwarfs with previously
published spectral types later than T8.  The classification of
\Wseventeenthirtyeight\ as a Y dwarf is bolstered by the distinctive
break in the trend of the NH$_3$-$H$ values with spectral type which
suggests that a new absorption band has indeed emerged at the T/Y dwarf
transition.  The remaining spectral indices do not show a break at the
T/Y transition, but the CH$_4$-$J$, H$_2$O-$H$, and W$_J$ indices do
show a smooth trend with spectral type down to Y0 indicating that they
can still be used as proxies for direct spectral comparisons.  Indeed
the value of the W$_J$ spectral index for \Wseventeenthirtyeight\ is far
from saturated so we support the suggestion by
\citet{2008MNRAS.391..320B} that this index can be used as a proxy for
direct comparison for late-type T dwarfs and early-type Y
dwarfs. However, the CH$_4$-$H$ index is clearly beginning to saturate
at T9 and the H$_2$O-$J$ index may even reverse at Y0 rendering these
indices less useful for classification purposes.  Finally, although
there is scatter due to the very low S/N of some of the spectra, the new
WISE brown dwarfs are clearly distinct from the previous $\ge$T8.5
dwarfs and cluster around the Y0 spectral standard.

\subsection{Atmospheric and Structural Properties}

\subsubsection{Atmospheric Properties}

\label{sec:atmosprop}

In order to investigate the atmospheric properties (e.g., \teff, \logg)
of the brown dwarfs, we have compared their near-infrared spectra to a
new preliminary grid of model spectra generated with the model
atmospheres of Marley \& Saumon.  A detailed description of the basic
models can be found in \citet{2002ApJ...568..335M},
\citet{2008ApJ...689.1327S}, \citet{2008ApJ...678.1372C}, and
\citet{2009ApJ...702..154S}.  This preliminary grid includes a new
NH$_3$ line list \citep{2011MNRAS.413.1828Y} and a new prescription for
the collision induced opacity for H$_2$ (Saumon et al., in prep).  A
more detailed study that compares the model spectra to the near-infrared
spectra, and WISE and \textit{Spitzer} photometry is in preparation.

The grid consists of solar metallicity, cloudless models with the
following parameters: \teff=200$-$1000 K in steps of 50 K; \logg\ =
3.75$-$ 5 in steps of 0.25 (cm s$^{-2}$); and \kzz=0, 10$^4$ cm$^2$
s$^{-1}$.  Although the opacities of the condensate clouds are not
included in the atmospheric models, i.e. they are cloudless, the effects
on the atmospheric chemistry due to the rainout of the condensates is
accounted for in the models.  This assumption is reasonable for the
silicate and liquid iron clouds since they form well below the
observable photosphere \citep[see however,][]{2010ApJ...725.1405B} but
may not be valid if, as expected, H$_2$O clouds form high in the
atmosphere of the coldest models.  The eddy diffusion coefficient, \kzz,
parametrizes the vigor of mixing in the radiative layers of the
atmosphere.  A value of \kzz$>$0 cm$^2$ s$^{-1}$ results in mixing that
can prevent the abundances of CO and CH$_4$ (the dominant carbon-bearing
species) from coming into chemical equilibrium because the mixing time
scales become shorter than the timescales of key chemical reactions
involved in the conversion of CO to CH$_4$
\citep{2002Icar..155..393L,2003IAUS..211..345S,2007ApJ...669.1248H}.
Typical values of \kzz\ in the stratospheres of giant planets are 10$^2$
to 10$^5$ cm$^2$ s$^{-1}$ \citep{2006ApJ...647..552S}.  The abundances
of N$_2$ and NH$_3$ (the dominant nitrogen-bearing species) are also
kept from coming into chemical equilibrium by mixing, but in this case
the mixing timescales are set in the convective layers of the atmosphere
by the mixing length theory.  As a result, the final non-equilibrium
abundances of N$_2$ and NH$_3$ are not sensitive to variations in the
eddy diffusion coefficient \kzz.  However by convention, models with
\kzz=0 cm$^2$ s$^{-1}$ are in full chemical equilibrium (i.e. the effect
of convective mixing on the nitrogen chemistry is ignored) and models
with \kzz $\ne$ 0 cm$^2$ s$^{-1}$ exhibit both carbon and nitrogen
non-equilibrium chemistry.

The best fitting models are identified using the goodness-of-fit
statistic, $G_k= \sum_{i=1}^n w_i \left ( \frac{f_i -
    C_k\mathcal{F}_{k,i}}{\sigma_i} \right)^2$, where $n$ is the number
of data pixels; $w_i$ is the weight for the $i$th wavelength (set to
unity in this case); $f_i$ and $F_{k,i}$ are the flux densities of the
data and model $k$, respectively; $\sigma_i$ are the errors in the
observed flux densities; and $C_k$ is an unknown multiplicative constant
equal to $(R/d)^2$, where $R$ is the radius of the star and $d$ is the
distance to the star \citep{2008ApJ...678.1372C}.  In order to increase
the S/N of the data, we first smoothed the higher resolution spectra to
$R$=200.  The model spectra were also smoothed to the same resolving
power and linearly interpolated onto the wavelength scale of the data.
For each model we compute the scale factor $C_k$ by minimizing $G_k$
with respect to $C_k$ and identify the best fitting model as having the
global minimum $G_k$ value.  To estimate the range of models that fits
the data well, we run a Monte Carlo simulation that uses the
uncertainties in the individual spectral points and the uncertainties in
the absolute flux calibration of the spectra to generate 10$^4$
simulated noisy spectra.  The fitting process is repeated on each
simulated spectrum and models that are consistent with the best fitting
model at the 3$\sigma$ level are considered equally good representations
of the data.  We did not attempt to fit the spectrum of
\Weighteentwentyeight\ because it is contaminated with light from other
stars in the WFC3 field of view (see \S\ref{sec:wfc3} and
\S\ref{sec:1828}).  After discussing the results of the fits to the
spectra of the other brown dwarfs, we return to estimate an approximate
effective temperature for \Weighteentwentyeight.

Table \ref{tab:properties} lists the best fitting model parameters for
each brown dwarf, as well as \UGPSzeroseventwotwo.  The derived
effective temperatures of the WISE brown dwarfs are all cold, ranging
from 350 to 500 K.  Indeed all but one have estimated effective
temperatures of less than 450 K making them the coldest
spectroscopically confirmed brown dwarfs known.  Five out of the six
best fitting models also have \kzz$\ne$0 which indicates that vertical
mixing is present in the atmospheres of these cold brown dwarfs.  This
is not a surprising result given that strong evidence for vertical
mixing in the atmospheres of brown dwarfs has been found at longer
wavelengths
\citep{2006ApJ...647..552S,2007ApJ...656.1136S,2007ApJ...655.1079L,2008ApJ...689L..53B,2009ApJ...695.1517L,2009ApJ...702..154S,2009ApJ...695..844G,2010AJ....140.1428C}.
At such low temperatures, the effects of non-equilibrium chemistry on
the $J$- and $H$-band spectra of brown dwarfs is limited to weakening
the NH$_3$ absorption bands.  The detection of mixing therefore
underscores the fact that NH$_3$ is probably at least partially
responsible for the absorption seen on the blue wing of the $H$-band
peak of \Wfourteenzerofive\ and \Wseventeenthirtyeight.

Figure \ref{fig:modcomp} shows the best fitting model spectra
overplotted on the data.  Since this is the first time such cold model
spectra have been compared to observed spectra, the agreement between
the models and the data is encouraging.  In particular, the height and
width of the $J$-band peaks are well matched by the model spectra.
Previous studies of hotter brown dwarfs fail to match both the peak and
width of this peak
\citep[e.g.,][]{2009ApJ...695.1517L,2011ApJ...735..116B}.  The improved
fits may be a result of the fact that we are fitting a smaller
wavelength range \citep{2008ApJ...678.1372C,2009AIPC.1094..283S} and/or
because the high $J$ vibration-rotation lines (the so-called ``hot''
lines), whose cross-sections are less well known, become less important
at such cold temperatures.

The models do, however, provide a poor fit to the blue wing of the
$H$-band peak of the spectra and fail to reproduce the heights of the
$Y$-band peaks of \Wzerofourten\ and \Wfifteenfortyone.  Note that the
peak of the $Y$-band is shaped by the 2$\nu_1$$+$2$\nu_4$ band of NH$_3$
centered at about 1.03 $\mu$m and therefore $Y$-band spectra of cold
brown dwarfs could provide the first clear detection of NH$_3$ at
near-infrared wavelengths.  In principle, the blue wing of the $H$-band
model spectrum could be brought into better agreement with the data by
further reducing the abundance of NH$_3$.  However, as noted above, the
abundance of NH$_3$ is insensitive to variations in \kzz\ because it is
quenched in the convective region where the mixing time scale is set by
the mixing length theory and not by the eddy diffusion coefficient.
Therefore, the mismatch between the data and models is most likely a
result of some other inadequacy in the model atmospheres.

Finally, although we cannot fit the models to the spectrum of
\Weighteentwentyeight, we can still estimate a rough effective
temperature.  The most salient feature of the spectrum is that the $J$-
and $H$-band peaks are roughly the same height in flux density units of
$f_\lambda$.  Only model spectra with \teff\ $\leq$ 250 K have $J$-band
peak fluxes that are equal to or less than the $H$-band peak fluxes.  A
second estimate of the effective temperature can be derived from the
observed $J-$W2 color of 9.29$\pm$0.35. We computed synthetic Mauna Kea
Observatories Near-Infrared \citep[MKO-NIR,][]{2002PASP..114..180T} $J$
and W2 magnitudes for each model in the grid and find that model spectra
with \teff=275$-$300 K have $J-$W2 colors that fall within
$\pm$2$\sigma$ of the observed color.  Taken together, these estimates
suggest that an appropriate upper limit to the effective temperature of
\Weighteentwentyeight\ is $\sim$300 K which makes \Weighteentwentyeight\
the coolest spectroscopically confirmed brown dwarf known.

\subsubsection{Structural Properties}

\label{sec:strucprop}

With estimates of the effective temperatures and surface gravities of
the new brown dwarfs in hand, we can also estimate their radii ($R$) and
masses ($M$) using evolutionary models.  We used the cloudless structure
models of \citet{2008ApJ...689.1327S} because they used atmospheric
models that are nearly identical to the ones we used in our analysis for
boundary conditions.  As a result, the derived \teff, \logg, $R$, and
$M$ estimates are all self-consistent.  The radii and masses of the
brown dwarfs are given in Table \ref{tab:properties}.

\subsection{Spectroscopic Distance Estimates}

The value of the multiplicative constant $C_k$=$(R/d)^2$ derived as a by
product of the atmospheric model fitting procedure can be used to
estimate a so-called ``spectroscopic distance'' ($d_\mathrm{spec}$) to
brown dwarfs if their radii can be determined
\citep[e.g.,][]{2009ApJ...706.1114B}. In the absence of direct
measurements of brown dwarf radii, we can use the radii computed using
evolutionary models and (\teff, \logg) values.  The spectroscopic
distances of the new WISE brown dwarfs and \UGPSzeroseventwotwo\ are
given in Table \ref{tab:dist}.  Also listed in Table \ref{tab:dist} are
the photometric distance estimates of the WISE brown dwarfs from
\citet{Davy11} and parallactic distances to \UGPSzeroseventwotwo\
\citep{2010MNRAS.408L..56L} and \Wfifteenfortyone\ \citep{Davy11}.  The
former distance estimates are computed using a spectral type-W2 relation
derived from known brown dwarfs with spectral types ranging from L0 to
T9 and with $\pi / \sigma_\pi$ $>$ 3.  The photometric distances of the
new WISE brown dwarfs are based on an \textit{extrapolation} of this
relation and therefore should be treated with caution.

The agreement between the three distance estimates range from good to
poor.  For example, the spectroscopic and parallactic distances of
\Wfifteenfortyone\ are in good agreement but the photometric distance is
discrepant by a factor of two to four.  Perhaps most discouraging is the
mismatch between the spectroscopic and parallactic distances of
\UGPSzeroseventwotwo.  \citet{2011arXiv1103.0014L} recently showed that
the agreement between the spectroscopic distances (derived using only
near-infrared spectra) and the parallactic distances of ten late-type T
dwarfs range from 10\% to a factor of two, with no apparent trend with
spectral type or distance.  This suggests that spectroscopic distances
should only be used to confirm that the physical properties of brown
dwarfs (\teff, $R$, $M$) derived from atmospheric and evolutionary
models are consistent with the known distance to the brown dwarf.

A corollary to this statement is that if the spectroscopic and
parallactic distances are discrepant then some combination of the \teff,
\logg\ and $R$ values are in error.  In order to estimate the
significance of the bias in the spectroscopic distance estimate
introduced by systematic errors in the inferred atmospheric properties,
we have run a 1$-$2.5 $\mu$m model with \teff=600 K, \logg=4.5 cm
s$^{-2}$ through the fitting procedure described in
\S\ref{sec:atmosprop}.  The model spectrum was first multiplied by an
appropriate value of ($R/d$)$^2$ corresponding to 10 pc.  Figure
\ref{fig:distcomp} shows the ratio of $d_\mathrm{spec}$/ 10 pc derived
for models with effective temperatures from 500 to 700 K and surface
gravities from 3.75 to 5.0.  The maximum change in $d_\mathrm{spec}$ for
these models is approximately a factor of two for a change of $+$100
K/$-$0.75 dex and $-$100 K/$+$0.5 dex in \teff/\logg.  The spectroscopic
distance is also most sensitive to changes in \teff\ as noted by
\citet{2011arXiv1103.0014L}.

The apparent mismatch between the spectroscopic and parallactic distance
estimates is perhaps not so surprising as \citet{2008ApJ...678.1372C}
have shown that variations of order 100 to 200 K are common when
estimating the effective temperatures of L and early- to mid-type T
dwarfs using spectra that cover only a fraction of the spectral energy
distribution. These variations are most likely exacerbated by the fact
that only $\sim$35\% (in flux density units of f$_\lambda$) of the
bolometric flux of a \teff=600 K brown dwarf is emitted at near-infrared
wavelengths.  In summary, given the uncertainties in the model
atmospheres and the difficulty in estimating the effective temperatures
and surface gravities of brown dwarfs, it is not surprising that
spectroscopic distance estimates do not always agree with the
parallactic distances.

\section{Discussion}

The new WISE brown dwarfs presented herein are the coldest
(\teff=300$-$500 K) spectroscopically confirmed brown dwarfs currently
known.  However as noted in \S1, \WDzeroeightzerosixFull B and
\CFHfourteenfiftyeight B have estimated effective temperatures of
$\sim$300$-$400 K based on photometry alone
\citep{2011ApJ...730L...9L,2011arXiv1103.0014L,2011ApJ...732L..29R}.
How do the properties of these two brown dwarfs compare with the new
WISE brown dwarfs?

The upper panel of Figure \ref{fig:bdcomp} shows the absolute $J$-band
magnitude (M$_J$) as a function of spectral type for a sample of field T
dwarfs \citep{2010ApJ...710.1627L}, \Wfifteenfortyone\ (the only WISE
brown dwarf with a measured parallax), \WDzeroeightzerosixFull B (Luhman
et al. submitted) and \CFHfourteenfiftyeight B
\citep{2011arXiv1103.0014L}.  The value of M$_J$ increases precipitously
beyond a spectral type of T8 and reaches $\sim$23.9 for
\Wfifteenfortyone\ (Y0).  The absolute magnitude of
\CFHfourteenfiftyeight B falls between the two T9 dwarfs and
\Wfifteenfortyone\ suggesting that it has a spectral type of T9$-$Y0.
However the absolute magnitudes of more late-type T dwarfs and Y dwarfs
must be measured before any strong conclusions can be drawn based on
absolute magnitudes alone.  The absolute magnitude of
\WDzeroeightzerosixFull B is still only a limit (M$_J$ $>$ 22.5) which
leaves open the possibility that \WDzeroeightzerosixFull B is even
fainter, and thus presumably cooler than, \Wfifteenfortyone.  Either
way, it is clear that based on the trend of M$_J$ with spectral type
that observing even colder objects at near-infrared wavelengths is going
to become increasingly difficult unless they are very close the Sun.

The lower panel of Figure \ref{fig:bdcomp} shows the $J-H$ colors as a
function of spectral type for a sample of field T dwarfs
\citep{2010ApJ...710.1627L}, the new WISE brown dwarfs, and
\CFHfourteenfiftyeight B (\WDzeroeightzerosixFull B has yet to be
detected in either the $J$ or $H$ band).  Some of the colors of the WISE
brown dwarfs have large uncertainties so we also computed synthetic
MKO-NIR $J-H$ colors as described in \citet{2009ApJS..185..289R}; they
are shown as triangles in Figure \ref{fig:bdcomp}.  Since the
WFC3/\textit{HST} spectra do not span the entire $H$ band, we used the
spectrum of \UGPSzeroseventwotwo\ to extend the spectra of
\Wseventeenthirtyeight\ and \Wfourteenzerofive\ to the limit of the $H$
band filter.  The synthetic and observed photometry of
\Wfourteenzerofive\ are clearly discrepant and it is unclear what the
underlying cause is.  The scatter in the $J-H$ colors of the dwarfs at
the T/Y transition makes it difficult to assign a spectral type to
\CFHfourteenfiftyeight B but it is broadly consistent with T6$-$Y0.  The
overall trend with spectral type suggests suggests that the $J-H$ colors
may be turning towards to the red at the T/Y transition.  This turn is
broadly consistent with theoretical models which predict that the $J-K$
color also turns towards the red at \teff=390$-$450 K
\citep{2003ApJ...596..587B}.  However given the small number of objects
and the large uncertainties in colors, a definitive conclusion can not
yet be reached.

Finally, given the rapid increase in the absolute $J$-band magnitude at
the T/Y transition, it is reasonable to ask whether the near-infrared is
the appropriate wavelength range with which to define the Y spectral
class.  Indeed historically as cooler and cooler stars were discovered,
the wavelength range used for spectral classification moved from the
blue violet at 3900$-$4900 \AA\ \citep{1943QB881.M6.......}, through the
red optical at 7000$-$10000 \AA\
\citep{1976PhDT........14B,1991ApJS...77..417K,1999ApJ...519..802K}, and
finally into the near infrared at 1$-$2.5 $\mu$m
\citep{2006ApJ...637.1067B}.  Since Y dwarfs emit the majority of their
flux in the mid-infrared, it seems only natural to devise the spectral
classification system at these wavelengths.  The smooth spectral
morphological changes seen in the 5.5$-$14.5 $\mu$m \textit{Spitzer}
spectra of L and T dwarfs \citep{2006ApJ...648..614C} suggest that a
mid-infrared classification scheme for the Y dwarfs is plausible.
Unfortunately, observing at wavelengths longer about 2.5 $\mu$m from the
ground is exceedingly difficult due to the high thermal background.
Observations conducted from space do not suffer from this limitation but
there are currently no space-based facilities (including
\textit{Spitzer}) capable of obtaining mid-infrared spectra of cold
brown dwarfs.  The \textit{James Webb Space Telescope} will provide a
platform with which to observe cold brown dwarfs
\citep{2003ApJ...596..587B,2009and..book..101M} but its launch is, at
best, still years away.

We are therefore left in the unfortunate position of either waiting for
a (space- or ground-based) facility capable of sensitive mid-infrared
observations or constructing a classification scheme in the near
infrared.  Given the large number of cold brown dwarfs now known, we
believe it is important to devise a system with which to classify them
based on the data currently available.  The creation of a near-infrared
scheme in no way invalidates any future mid-infrared system that may be
devised.  Indeed the classification of brown dwarfs at two different
wavelengths is not unprecedented as both the L and the T dwarfs have
classification systems based in the red optical
\citep{1999ApJ...519..802K,2003ApJ...594..510B} and the near infrared
\citep{2002ApJ...564..466G,2010ApJS..190..100K,2006ApJ...637.1067B}.
Ultimately, the utility of any classification system will be measured by
whether or not it is adopted by the brown dwarf community.

\section{Summary}

As part of an ongoing search for the coldest brown dwarfs in the solar
neighborhood using the Wide-field Infrared Survey Explorer, we have
discovered seven ultracool brown dwarfs whose near-infrared spectra
indicate that they are the latest-type brown dwarfs currently known.
Based on the spectral morphological differences between these brown
dwarfs and the late-type T dwarfs, we have identified six of them as the
first members of the Y spectral class.  A comparison to the model
spectra of Marley \& Saumon indicate that they have effective
temperatures ranging from 300 to 500 K and thus are the coolest
spectroscopically confirmed brown dwarfs currently known.

\clearpage

\acknowledgements

We thank Tom Jarrett for guidance with the WIRC data reduction, Barry
Rothberg and Norbert Pirzkal for their guidance in reducing the HST/WFC3
data, and Ben Burningham, Sandy Leggett, and Mike Liu for providing
digital copies of late-type T dwarf spectra.  We also thank Mauricio
Martinez, Jorge Araya, and Nidia Morrell for observing support at
Magellan.  M.S.M. and D.S. acknowledge the support of the NASA ATP
program.  This publication makes use of data products from the
Wide-field Infrared Survey Explorer (WISE), the Two Micron All Sky
Survey (2MASS) and the Sloan Digital Sky Survey (SDSS).  WISE is a joint
project of the University of California, Los Angeles, and the Jet
Propulsion Laboratory/California Institute of Technology, funded by the
National Aeronautics and Space Administration.  2MASS is a joint project
of the University of Massachusetts and the Infrared Processing and
Analysis Center/California Institute of Technology, funded by the
National Aeronautics and Space Administration and the National Science
Foundation. SDSS is funded by the Alfred P. Sloan Foundation, the
Participating Institutions, the National Science Foundation, the
U.S. Department of Energy, the National Aeronautics and Space
Administration, the Japanese Monbukagakusho, the Max Planck Society, and
the Higher Education Funding Council for England.  This research has
made use of the NASA/IPAC Infrared Science Archive (IRSA), which is
operated by the Jet Propulsion Laboratory, California Institute of
Technology, under contract with the National Aeronautics and Space
Administration. Our research has also benefited from the M, L, and T
dwarf compendium housed at DwarfArchives.org whose server was funded by
a NASA Small Research Grant, administered by the American Astronomical
Society and the SpeX Prism Spectral Libraries, maintained by Adam
Burgasser at \url{http://www.browndwarfs.org/spexprism}. Data presented
herein were obtained at the W. M. Keck Observatory from telescope time
allocated to the National Aeronautics and Space Administration through
the agency's scientific partnership with the California Institute of
Technology and the University of California.  The Observatory was made
possible by the generous financial support of the W. M. Keck Foundation.
This paper also includes data gathered with the 6.5 m Magellan
Telescopes located at Las Campanas Observatory, Chile and the Peters
Automated Infrared Imaging Telescope (PAIRITEL) which is operated by the
Smithsonian Astrophysical Observatory (SAO) and was made possible by a
grant from the Harvard University Milton Fund, the camera loan from the
University of Virginia, and the continued support of the SAO and the
University of California, Berkeley. Partial support for PAIRITEL
operations and this work comes from National Aeronautics and Space
Administration grant NNG06GH50G.  AJB acknowledges support from the
Chris and Warren Hellman Fellowship Program.  Finally, this research was
supported (in part) by an appointment to the NASA Postdoctoral Program
at the Jet Propulsion Laboratory, administered by Oak Ridge Associated
Universities through a contract with NASA.

{\it Facilities:} \facility{IRTF (SpeX)}, \facility{Palomar (WIRC)}, \facility{Magellan (FIRE)}, \facility{Magellan (PANIC)}, \facility{Keck (NIRC2)}, \facility{\textit{Spitzer} (IRAC)}, \facility{Keck (NIRSPEC)}, \facility{\textit{HST} (WFC3)}

\clearpage

\bibliographystyle{apj}
\bibliography{tmp,ref2}

\begin{thebibliography}{93}
\expandafter\ifx\csname natexlab\endcsname\relax\def\natexlab#1{#1}\fi

\bibitem[{{Allen} {et~al.}(2005){Allen}, {Koerner}, {Reid}, \&
  {Trilling}}]{2005ApJ...625..385A}
{Allen}, P.~R., {Koerner}, D.~W., {Reid}, I.~N., \& {Trilling}, D.~E. 2005,
  \apj, 625, 385

\bibitem[{{Bertin} \& {Arnouts}(1996)}]{1996A&AS..117..393B}
{Bertin}, E., \& {Arnouts}, S. 1996, \aaps, 117, 393

\bibitem[{{Beuzit} {et~al.}(2006){Beuzit}, {Mouillet}, {Moutou}, {Dohlen},
  {Fusco}, {Puget}, {Udry}, {Gratton}, {Schmid}, {Feldt}, {Kasper}, \& {The
  Vlt-Pf Consortium}}]{2006tafp.conf..353B}
{Beuzit}, J.~L., {et~al.} 2006, in Tenth Anniversary of 51 Peg-b: Status of and
  prospects for hot Jupiter studies, ed. {L.~Arnold, F.~Bouchy, \& C.~Moutou},
  353--355

\bibitem[{{Boeshaar}(1976)}]{1976PhDT........14B}
{Boeshaar}, P.~C. 1976, PhD thesis, Ohio State University, Columbus.

\bibitem[{{Bowler} {et~al.}(2009){Bowler}, {Liu}, \&
  {Cushing}}]{2009ApJ...706.1114B}
{Bowler}, B.~P., {Liu}, M.~C., \& {Cushing}, M.~C. 2009, \apj, 706, 1114

\bibitem[{{Burgasser}(2004)}]{2004ApJS..155..191B}
{Burgasser}, A.~J. 2004, \apjs, 155, 191

\bibitem[{{Burgasser} {et~al.}(2011){Burgasser}, {Cushing}, {Kirkpatrick},
  {Gelino}, {Griffith}, {Looper}, {Tinney}, {Simcoe}, {Bochanski}, {Skrutskie},
  {Mainzer}, {Thompson}, {Marsh}, {Bauer}, \& {Wright}}]{2011ApJ...735..116B}
{Burgasser}, A.~J., {et~al.} 2011, \apj, 735, 116

\bibitem[{{Burgasser} {et~al.}(2006){Burgasser}, {Geballe}, {Leggett},
  {Kirkpatrick}, \& {Golimowski}}]{2006ApJ...637.1067B}
{Burgasser}, A.~J., {Geballe}, T.~R., {Leggett}, S.~K., {Kirkpatrick}, J.~D.,
  \& {Golimowski}, D.~A. 2006, \apj, 637, 1067

\bibitem[{{Burgasser} {et~al.}(2003){Burgasser}, {Kirkpatrick}, {Liebert}, \&
  {Burrows}}]{2003ApJ...594..510B}
{Burgasser}, A.~J., {Kirkpatrick}, J.~D., {Liebert}, J., \& {Burrows}, A. 2003,
  \apj, 594, 510

\bibitem[{{Burgasser} {et~al.}(2010){Burgasser}, {Simcoe}, {Bochanski},
  {Saumon}, {Mamajek}, {Cushing}, {Marley}, {McMurtry}, {Pipher}, \&
  {Forrest}}]{2010ApJ...725.1405B}
{Burgasser}, A.~J., {et~al.} 2010, \apj, 725, 1405

\bibitem[{{Burgasser} {et~al.}(2008){Burgasser}, {Tinney}, {Cushing}, {Saumon},
  {Marley}, {Bennett}, \& {Kirkpatrick}}]{2008ApJ...689L..53B}
{Burgasser}, A.~J., {Tinney}, C.~G., {Cushing}, M.~C., {Saumon}, D., {Marley},
  M.~S., {Bennett}, C.~S., \& {Kirkpatrick}, J.~D. 2008, \apjl, 689, L53

\bibitem[{{Burningham} {et~al.}(2011{\natexlab{a}}){Burningham}, {Leggett},
  {Homeier}, {Saumon}, {Lucas}, {Pinfield}, {Tinney}, {Allard}, {Marley},
  {Jones}, {Murray}, {Ishii}, {Day-Jones}, {Gomes}, \&
  {Zhang}}]{2011MNRAS.414.3590B}
{Burningham}, B., {et~al.} 2011{\natexlab{a}}, \mnras, 414, 3590

\bibitem[{{Burningham} {et~al.}(2011{\natexlab{b}}){Burningham}, {Lucas},
  {Leggett}, {Smart}, {Baker}, {Pinfield}, {Tinney}, {Homeier}, {Allard},
  {Zhang}, {Gomes}, {Day-Jones}, {Jones}, {Kov{\'a}cs}, {Lodieu}, {Marocco},
  {Murray}, \& {Sip{\H o}cz}}]{2011MNRAS.414L..90B}
---. 2011{\natexlab{b}}, \mnras, 414, L90

\bibitem[{{Burningham} {et~al.}(2008){Burningham}, {Pinfield}, {Leggett},
  {Tamura}, {Lucas}, {Homeier}, {Day-Jones}, {Jones}, {Clarke}, {Ishii},
  {Kuzuhara}, {Lodieu}, {Zapatero Osorio}, {Venemans}, {Mortlock}, {Barrado Y
  Navascu{\'e}s}, {Martin}, \& {Magazz{\`u}}}]{2008MNRAS.391..320B}
---. 2008, \mnras, 391, 320

\bibitem[{{Burningham} {et~al.}(2009){Burningham}, {Pinfield}, {Leggett},
  {Tinney}, {Liu}, {Homeier}, {West}, {Day-Jones}, {Huelamo}, {Dupuy}, {Zhang},
  {Murray}, {Lodieu}, {Barrado Y Navascu{\'e}s}, {Folkes}, {Galvez-Ortiz},
  {Jones}, {Lucas}, {Calderon}, \& {Tamura}}]{2009MNRAS.395.1237B}
---. 2009, \mnras, 395, 1237

\bibitem[{{Burningham} {et~al.}(2010){Burningham}, {Pinfield}, {Lucas},
  {Leggett}, {Deacon}, {Tamura}, {Tinney}, {Lodieu}, {Zhang}, {Huelamo},
  {Jones}, {Murray}, {Mortlock}, {Patel}, {Barrado Y Navascu{\'e}s}, {Zapatero
  Osorio}, {Ishii}, {Kuzuhara}, \& {Smart}}]{2010MNRAS.406.1885B}
---. 2010, \mnras, 406, 1885

\bibitem[{{Burrows} \& {Sharp}(1999)}]{1999ApJ...512..843B}
{Burrows}, A., \& {Sharp}, C.~M. 1999, \apj, 512, 843

\bibitem[{{Burrows} {et~al.}(2003){Burrows}, {Sudarsky}, \&
  {Lunine}}]{2003ApJ...596..587B}
{Burrows}, A., {Sudarsky}, D., \& {Lunine}, J.~I. 2003, \apj, 596, 587

\bibitem[{{Chauvin} {et~al.}(2004){Chauvin}, {Lagrange}, {Dumas}, {Zuckerman},
  {Mouillet}, {Song}, {Beuzit}, \& {Lowrance}}]{2004A&A...425L..29C}
{Chauvin}, G., {Lagrange}, A., {Dumas}, C., {Zuckerman}, B., {Mouillet}, D.,
  {Song}, I., {Beuzit}, J., \& {Lowrance}, P. 2004, \aap, 425, L29

\bibitem[{{Cushing} {et~al.}(2008){Cushing}, {Marley}, {Saumon}, {Kelly},
  {Vacca}, {Rayner}, {Freedman}, {Lodders}, \& {Roellig}}]{2008ApJ...678.1372C}
{Cushing}, M.~C., {et~al.} 2008, \apj, 678, 1372

\bibitem[{{Cushing} {et~al.}(2005){Cushing}, {Rayner}, \&
  {Vacca}}]{2005ApJ...623.1115C}
{Cushing}, M.~C., {Rayner}, J.~T., \& {Vacca}, W.~D. 2005, \apj, 623, 1115

\bibitem[{{Cushing} {et~al.}(2006){Cushing}, {Roellig}, {Marley}, {Saumon},
  {Leggett}, {Kirkpatrick}, {Wilson}, {Sloan}, {Mainzer}, {Van Cleve}, \&
  {Houck}}]{2006ApJ...648..614C}
{Cushing}, M.~C., {et~al.} 2006, \apj, 648, 614

\bibitem[{{Cushing} {et~al.}(2010){Cushing}, {Saumon}, \&
  {Marley}}]{2010AJ....140.1428C}
{Cushing}, M.~C., {Saumon}, D., \& {Marley}, M.~S. 2010, \aj, 140, 1428

\bibitem[{{Delorme} {et~al.}(2008{\natexlab{a}}){Delorme}, {Delfosse},
  {Albert}, {Artigau}, {Forveille}, {Reyl{\'e}}, {Allard}, {Homeier}, {Robin},
  {Willott}, {Liu}, \& {Dupuy}}]{2008A&A...482..961D}
{Delorme}, P., {et~al.} 2008{\natexlab{a}}, \aap, 482, 961

\bibitem[{{Delorme} {et~al.}(2008{\natexlab{b}}){Delorme}, {Willott},
  {Forveille}, {Delfosse}, {Reyl{\'e}}, {Bertin}, {Albert}, {Artigau}, {Robin},
  {Allard}, {Doyon}, \& {Hill}}]{2008A&A...484..469D}
---. 2008{\natexlab{b}}, \aap, 484, 469

\bibitem[{{Epchtein} {et~al.}(1997){Epchtein}, {de Batz}, {Capoani},
  {Chevallier}, {Copet}, {Fouqu{\'e}}, {Lacombe}, {Le Bertre}, {Pau}, {Rouan},
  {Ruphy}, {Simon}, {Tiph{\`e}ne}, {Burton}, {Bertin}, {Deul}, {Habing},
  {Borsenberger}, {Dennefeld}, {Guglielmo}, {Loup}, {Mamon}, {Ng}, {Omont},
  {Provost}, {Renault}, {Tanguy}, {Kimeswenger}, {Kienel}, {Garzon}, {Persi},
  {Ferrari-Toniolo}, {Robin}, {Paturel}, {Vauglin}, {Forveille}, {Delfosse},
  {Hron}, {Schultheis}, {Appenzeller}, {Wagner}, {Balazs}, {Holl},
  {L{\'e}pine}, {Boscolo}, {Picazzio}, {Duc}, \&
  {Mennessier}}]{1997Msngr..87...27E}
{Epchtein}, N., {et~al.} 1997, The Messenger, 87, 27

\bibitem[{{Fazio} {et~al.}(2004){Fazio}, {Hora}, {Allen}, {Ashby}, {Barmby},
  {Deutsch}, {Huang}, {Kleiner}, {Marengo}, {Megeath}, {Melnick}, {Pahre},
  {Patten}, {Polizotti}, {Smith}, {Taylor}, {Wang}, {Willner}, {Hoffmann},
  {Pipher}, {Forrest}, {McMurty}, {McCreight}, {McKelvey}, {McMurray}, {Koch},
  {Moseley}, {Arendt}, {Mentzell}, {Marx}, {Losch}, {Mayman}, {Eichhorn},
  {Krebs}, {Jhabvala}, {Gezari}, {Fixsen}, {Flores}, {Shakoorzadeh}, {Jungo},
  {Hakun}, {Workman}, {Karpati}, {Kichak}, {Whitley}, {Mann}, {Tollestrup},
  {Eisenhardt}, {Stern}, {Gorjian}, {Bhattacharya}, {Carey}, {Nelson},
  {Glaccum}, {Lacy}, {Lowrance}, {Laine}, {Reach}, {Stauffer}, {Surace},
  {Wilson}, {Wright}, {Hoffman}, {Domingo}, \& {Cohen}}]{2004ApJS..154...10F}
{Fazio}, G.~G., {et~al.} 2004, \apjs, 154, 10

\bibitem[{{Freedman} {et~al.}(2008){Freedman}, {Marley}, \&
  {Lodders}}]{2008ApJS..174..504F}
{Freedman}, R.~S., {Marley}, M.~S., \& {Lodders}, K. 2008, \apjs, 174, 504

\bibitem[{{Geballe} {et~al.}(2002){Geballe}, {Knapp}, {Leggett}, {Fan},
  {Golimowski}, {Anderson}, {Brinkmann}, {Csabai}, {Gunn}, {Hawley},
  {Hennessy}, {Henry}, {Hill}, {Hindsley}, {Ivezi{\'c}}, {Lupton}, {McDaniel},
  {Munn}, {Narayanan}, {Peng}, {Pier}, {Rockosi}, {Schneider}, {Smith},
  {Strauss}, {Tsvetanov}, {Uomoto}, {York}, \& {Zheng}}]{2002ApJ...564..466G}
{Geballe}, T.~R., {et~al.} 2002, \apj, 564, 466

\bibitem[{{Geballe} {et~al.}(2009){Geballe}, {Saumon}, {Golimowski}, {Leggett},
  {Marley}, \& {Noll}}]{2009ApJ...695..844G}
{Geballe}, T.~R., {Saumon}, D., {Golimowski}, D.~A., {Leggett}, S.~K.,
  {Marley}, M.~S., \& {Noll}, K.~S. 2009, \apj, 695, 844

\bibitem[{{Hanel} {et~al.}(1981){Hanel}, {Conrath}, {Herath}, {Kunde}, \&
  {Pirraglia}}]{1981JGR....86.8705H}
{Hanel}, R., {Conrath}, B., {Herath}, L., {Kunde}, V., \& {Pirraglia}, J. 1981,
  \jgr, 86, 8705

\bibitem[{{Hayashi} \& {Nakano}(1963)}]{1963PThPh..30..460H}
{Hayashi}, C., \& {Nakano}, T. 1963, Progress of Theoretical Physics, 30, 460

\bibitem[{{Hinkley} {et~al.}(2011){Hinkley}, {Oppenheimer}, {Zimmerman},
  {Brenner}, {Parry}, {Crepp}, {Vasisht}, {Ligon}, {King}, {Soummer},
  {Sivaramakrishnan}, {Beichman}, {Shao}, {Roberts}, {Bouchez}, {Dekany},
  {Pueyo}, {Roberts}, {Lockhart}, {Zhai}, {Shelton}, \&
  {Burruss}}]{2011PASP..123...74H}
{Hinkley}, S., {et~al.} 2011, \pasp, 123, 74

\bibitem[{{Hubeny} \& {Burrows}(2007)}]{2007ApJ...669.1248H}
{Hubeny}, I., \& {Burrows}, A. 2007, \apj, 669, 1248

\bibitem[{{Kimble} {et~al.}(2008){Kimble}, {MacKenty}, {O'Connell}, \&
  {Townsend}}]{2008SPIE.7010E..43K}
{Kimble}, R.~A., {MacKenty}, J.~W., {O'Connell}, R.~W., \& {Townsend}, J.~A.
  2008, in Presented at the Society of Photo-Optical Instrumentation Engineers
  (SPIE) Conference, Vol. 7010, Society of Photo-Optical Instrumentation
  Engineers (SPIE) Conference Series

\bibitem[{{Kirkpatrick}(2000)}]{2000ASPC..212...20K}
{Kirkpatrick}, J.~D. 2000, in Astronomical Society of the Pacific Conference
  Series, Vol. 212, From Giant Planets to Cool Stars, ed. {C.~A.~Griffith \&
  M.~S.~Marley}, 20

\bibitem[{{Kirkpatrick}(2008)}]{2008ASPC..384...85K}
{Kirkpatrick}, J.~D. 2008, in Astronomical Society of the Pacific Conference
  Series, Vol. 384, 14th Cambridge Workshop on Cool Stars, Stellar Systems, and
  the Sun, ed. {G.~van Belle}, 85

\bibitem[{{Kirkpatrick} {et~al.}(2011){Kirkpatrick}, {Cushing}, {Gelino},
  {Griffith}, {Skrutskie}, {Marsh}, {Wright}, {Mainzer}, {Eisenhardt},
  {McLean}, {Thompson}, {Bauer}, {Benford}, \& {Bridge}}]{Davy11}
{Kirkpatrick}, J.~D., {et~al.} 2011, \apjs

\bibitem[{{Kirkpatrick} {et~al.}(1991){Kirkpatrick}, {Henry}, \&
  {McCarthy}}]{1991ApJS...77..417K}
{Kirkpatrick}, J.~D., {Henry}, T.~J., \& {McCarthy}, Jr., D.~W. 1991, \apjs,
  77, 417

\bibitem[{{Kirkpatrick} {et~al.}(2010){Kirkpatrick}, {Looper}, {Burgasser},
  {Schurr}, {Cutri}, {Cushing}, {Cruz}, {Sweet}, {Knapp}, {Barman},
  {Bochanski}, {Roellig}, {McLean}, {McGovern}, \&
  {Rice}}]{2010ApJS..190..100K}
{Kirkpatrick}, J.~D., {et~al.} 2010, \apjs, 190, 100

\bibitem[{{Kirkpatrick} {et~al.}(1999){Kirkpatrick}, {Reid}, {Liebert},
  {Cutri}, {Nelson}, {Beichman}, {Dahn}, {Monet}, {Gizis}, \&
  {Skrutskie}}]{1999ApJ...519..802K}
---. 1999, \apj, 519, 802

\bibitem[{{Koekemoer} {et~al.}(2002){Koekemoer}, {Fruchter}, {Hook}, \&
  {Hack}}]{2002hstc.conf..337K}
{Koekemoer}, A.~M., {Fruchter}, A.~S., {Hook}, R.~N., \& {Hack}, W. 2002, in
  The 2002 HST Calibration Workshop : Hubble after the Installation of the ACS
  and the NICMOS Cooling System, ed. {S.~Arribas, A.~Koekemoer, \&
  B.~Whitmore}, 337

\bibitem[{{Kumar}(1963)}]{1963ApJ...137.1121K}
{Kumar}, S.~S. 1963, \apj, 137, 1121

\bibitem[{{K{\"u}mmel} {et~al.}(2009){K{\"u}mmel}, {Walsh}, {Pirzkal},
  {Kuntschner}, \& {Pasquali}}]{2009PASP..121...59K}
{K{\"u}mmel}, M., {Walsh}, J.~R., {Pirzkal}, N., {Kuntschner}, H., \&
  {Pasquali}, A. 2009, \pasp, 121, 59

\bibitem[{{Lawrence} {et~al.}(2007){Lawrence}, {Warren}, {Almaini}, {Edge},
  {Hambly}, {Jameson}, {Lucas}, {Casali}, {Adamson}, {Dye}, {Emerson},
  {Foucaud}, {Hewett}, {Hirst}, {Hodgkin}, {Irwin}, {Lodieu}, {McMahon},
  {Simpson}, {Smail}, {Mortlock}, \& {Folger}}]{2007MNRAS.379.1599L}
{Lawrence}, A., {et~al.} 2007, \mnras, 379, 1599

\bibitem[{{Leggett} {et~al.}(2010){Leggett}, {Burningham}, {Saumon}, {Marley},
  {Warren}, {Smart}, {Jones}, {Lucas}, {Pinfield}, \&
  {Tamura}}]{2010ApJ...710.1627L}
{Leggett}, S.~K., {et~al.} 2010, \apj, 710, 1627

\bibitem[{{Leggett} {et~al.}(2009){Leggett}, {Cushing}, {Saumon}, {Marley},
  {Roellig}, {Warren}, {Burningham}, {Jones}, {Kirkpatrick}, {Lodieu}, {Lucas},
  {Mainzer}, {Mart{\'{\i}}n}, {McCaughrean}, {Pinfield}, {Sloan}, {Smart},
  {Tamura}, \& {Van Cleve}}]{2009ApJ...695.1517L}
---. 2009, \apj, 695, 1517

\bibitem[{{Leggett} {et~al.}(2007{\natexlab{a}}){Leggett}, {Marley},
  {Freedman}, {Saumon}, {Liu}, {Geballe}, {Golimowski}, \&
  {Stephens}}]{2007ApJ...667..537L}
{Leggett}, S.~K., {Marley}, M.~S., {Freedman}, R., {Saumon}, D., {Liu}, M.~C.,
  {Geballe}, T.~R., {Golimowski}, D.~A., \& {Stephens}, D.~C.
  2007{\natexlab{a}}, \apj, 667, 537

\bibitem[{{Leggett} {et~al.}(2007{\natexlab{b}}){Leggett}, {Saumon}, {Marley},
  {Geballe}, {Golimowski}, {Stephens}, \& {Fan}}]{2007ApJ...655.1079L}
{Leggett}, S.~K., {Saumon}, D., {Marley}, M.~S., {Geballe}, T.~R.,
  {Golimowski}, D.~A., {Stephens}, D., \& {Fan}, X. 2007{\natexlab{b}}, \apj,
  655, 1079

\bibitem[{{Liu} {et~al.}(2011){Liu}, {Delorme}, {Dupuy}, {Bowler}, {Albert},
  {Artigau}, {Reyle}, {Forveille}, \& {Delfosse}}]{2011arXiv1103.0014L}
{Liu}, M.~C., {et~al.} 2011, ArXiv e-prints

\bibitem[{{Lodders}(1999)}]{1999ApJ...519..793L}
{Lodders}, K. 1999, \apj, 519, 793

\bibitem[{{Lodders} \& {Fegley}(2002)}]{2002Icar..155..393L}
{Lodders}, K., \& {Fegley}, B. 2002, Icarus, 155, 393

\bibitem[{{Lucas} {et~al.}(2010){Lucas}, {Tinney}, {Burningham}, {Leggett},
  {Pinfield}, {Smart}, {Jones}, {Marocco}, {Barber}, {Yurchenko}, {Tennyson},
  {Ishii}, {Tamura}, {Day-Jones}, {Adamson}, {Allard}, \&
  {Homeier}}]{2010MNRAS.408L..56L}
{Lucas}, P.~W., {et~al.} 2010, \mnras, 408, L56

\bibitem[{{Luhman} {et~al.}(2005){Luhman}, {Adame}, {D'Alessio}, {Calvet},
  {Hartmann}, {Megeath}, \& {Fazio}}]{2005ApJ...635L..93L}
{Luhman}, K.~L., {Adame}, L., {D'Alessio}, P., {Calvet}, N., {Hartmann}, L.,
  {Megeath}, S.~T., \& {Fazio}, G.~G. 2005, \apjl, 635, L93

\bibitem[{{Luhman} {et~al.}(2011){Luhman}, {Burgasser}, \&
  {Bochanski}}]{2011ApJ...730L...9L}
{Luhman}, K.~L., {Burgasser}, A.~J., \& {Bochanski}, J.~J. 2011, \apjl, 730, L9

\bibitem[{{Macintosh} {et~al.}(2006){Macintosh}, {Graham}, {Palmer}, {Doyon},
  {Gavel}, {Larkin}, {Oppenheimer}, {Saddlemyer}, {Wallace}, {Bauman}, {Evans},
  {Erikson}, {Morzinski}, {Phillion}, {Poyneer}, {Sivaramakrishnan}, {Soummer},
  {Thibault}, \& {Veran}}]{2006SPIE.6272E..18M}
{Macintosh}, B., {et~al.} 2006, in Presented at the Society of Photo-Optical
  Instrumentation Engineers (SPIE) Conference, Vol. 6272, Society of
  Photo-Optical Instrumentation Engineers (SPIE) Conference Series

\bibitem[{{Mainzer} {et~al.}(2011){Mainzer}, {Cushing}, {Skrutskie}, {Gelino},
  {Kirkpatrick}, {Jarrett}, {Masci}, {Marley}, {Saumon}, {Wright}, {Beaton},
  {Dietrich}, {Eisenhardt}, {Garnavich}, {Kuhn}, {Leisawitz}, {Marsh},
  {McLean}, {Padgett}, \& {Rueff}}]{2011ApJ...726...30M}
{Mainzer}, A., {et~al.} 2011, \apj, 726, 30

\bibitem[{{Marley} \& {Leggett}(2009)}]{2009and..book..101M}
{Marley}, M.~S., \& {Leggett}, S.~K. 2009, {The Future of Ultracool Dwarf
  Science with JWST}, ed. {Thronson, H.~A., Stiavelli, M., \& Tielens, A.}, 101

\bibitem[{{Marley} {et~al.}(2002){Marley}, {Seager}, {Saumon}, {Lodders},
  {Ackerman}, {Freedman}, \& {Fan}}]{2002ApJ...568..335M}
{Marley}, M.~S., {Seager}, S., {Saumon}, D., {Lodders}, K., {Ackerman}, A.~S.,
  {Freedman}, R.~S., \& {Fan}, X. 2002, \apj, 568, 335

\bibitem[{{Martini} {et~al.}(2004){Martini}, {Persson}, {Murphy}, {Birk},
  {Shectman}, {Gunnels}, \& {Koch}}]{2004SPIE.5492.1653M}
{Martini}, P., {Persson}, S.~E., {Murphy}, D.~C., {Birk}, C., {Shectman},
  S.~A., {Gunnels}, S.~M., \& {Koch}, E. 2004, in Society of Photo-Optical
  Instrumentation Engineers (SPIE) Conference Series, Vol. 5492, Society of
  Photo-Optical Instrumentation Engineers (SPIE) Conference Series, ed.
  {A.~F.~M.~Moorwood \& M.~Iye}, 1653--1660

\bibitem[{{McLean} {et~al.}(1998){McLean}, {Becklin}, {Bendiksen}, {Brims},
  {Canfield}, {Figer}, {Graham}, {Hare}, {Lacayanga}, {Larkin}, {Larson},
  {Levenson}, {Magnone}, {Teplitz}, \& {Wong}}]{1998SPIE.3354..566M}
{McLean}, I.~S., {et~al.} 1998, in Society of Photo-Optical Instrumentation
  Engineers (SPIE) Conference Series, Vol. 3354, Society of Photo-Optical
  Instrumentation Engineers (SPIE) Conference Series, ed. {A.~M.~Fowler},
  566--578

\bibitem[{{McLean} {et~al.}(2000){McLean}, {Graham}, {Becklin}, {Figer},
  {Larkin}, {Levenson}, \& {Teplitz}}]{2000SPIE.4008.1048M}
{McLean}, I.~S., {Graham}, J.~R., {Becklin}, E.~E., {Figer}, D.~F., {Larkin},
  J.~E., {Levenson}, N.~A., \& {Teplitz}, H.~I. 2000, in Society of
  Photo-Optical Instrumentation Engineers (SPIE) Conference Series, Vol. 4008,
  Society of Photo-Optical Instrumentation Engineers (SPIE) Conference Series,
  ed. {M.~Iye \& A.~F.~Moorwood}, 1048--1055

\bibitem[{{McLean} {et~al.}(2003){McLean}, {McGovern}, {Burgasser},
  {Kirkpatrick}, {Prato}, \& {Kim}}]{2003ApJ...596..561M}
{McLean}, I.~S., {McGovern}, M.~R., {Burgasser}, A.~J., {Kirkpatrick}, J.~D.,
  {Prato}, L., \& {Kim}, S.~S. 2003, \apj, 596, 561

\bibitem[{{Morgan} {et~al.}(1943){Morgan}, {Keenan}, \&
  {Kellman}}]{1943QB881.M6.......}
{Morgan}, W.~W., {Keenan}, P.~C., \& {Kellman}, E. 1943, {An atlas of stellar
  spectra, with an outline of spectral classification}, ed. {Morgan, W.~W.,
  Keenan, P.~C., \& Kellman, E.}

\bibitem[{{Rayner} {et~al.}(2009){Rayner}, {Cushing}, \&
  {Vacca}}]{2009ApJS..185..289R}
{Rayner}, J.~T., {Cushing}, M.~C., \& {Vacca}, W.~D. 2009, \apjs, 185, 289

\bibitem[{{Rayner} {et~al.}(2003){Rayner}, {Toomey}, {Onaka}, {Denault},
  {Stahlberger}, {Vacca}, {Cushing}, \& {Wang}}]{2003PASP..115..362R}
{Rayner}, J.~T., {Toomey}, D.~W., {Onaka}, P.~M., {Denault}, A.~J.,
  {Stahlberger}, W.~E., {Vacca}, W.~D., {Cushing}, M.~C., \& {Wang}, S. 2003,
  \pasp, 115, 362

\bibitem[{{Rodriguez} {et~al.}(2011){Rodriguez}, {Zuckerman}, {Melis}, \&
  {Song}}]{2011ApJ...732L..29R}
{Rodriguez}, D.~R., {Zuckerman}, B., {Melis}, C., \& {Song}, I. 2011, \apjl,
  732, L29

\bibitem[{{Roellig} {et~al.}(2004){Roellig}, {Van Cleve}, {Sloan}, {Wilson},
  {Saumon}, {Leggett}, {Marley}, {Cushing}, {Kirkpatrick}, {Mainzer}, \&
  {Houck}}]{2004ApJS..154..418R}
{Roellig}, T.~L., {et~al.} 2004, \apjs, 154, 418

\bibitem[{{Saumon} {et~al.}(2000){Saumon}, {Geballe}, {Leggett}, {Marley},
  {Freedman}, {Lodders}, {Fegley}, \& {Sengupta}}]{2000ApJ...541..374S}
{Saumon}, D., {Geballe}, T.~R., {Leggett}, S.~K., {Marley}, M.~S., {Freedman},
  R.~S., {Lodders}, K., {Fegley}, Jr., B., \& {Sengupta}, S.~K. 2000, \apj,
  541, 374

\bibitem[{{Saumon} \& {Marley}(2008)}]{2008ApJ...689.1327S}
{Saumon}, D., \& {Marley}, M.~S. 2008, \apj, 689, 1327

\bibitem[{{Saumon} {et~al.}(2006){Saumon}, {Marley}, {Cushing}, {Leggett},
  {Roellig}, {Lodders}, \& {Freedman}}]{2006ApJ...647..552S}
{Saumon}, D., {Marley}, M.~S., {Cushing}, M.~C., {Leggett}, S.~K., {Roellig},
  T.~L., {Lodders}, K., \& {Freedman}, R.~S. 2006, \apj, 647, 552

\bibitem[{{Saumon} {et~al.}(2007){Saumon}, {Marley}, {Leggett}, {Geballe},
  {Stephens}, {Golimowski}, {Cushing}, {Fan}, {Rayner}, {Lodders}, \&
  {Freedman}}]{2007ApJ...656.1136S}
{Saumon}, D., {et~al.} 2007, \apj, 656, 1136

\bibitem[{{Saumon} {et~al.}(2003){Saumon}, {Marley}, {Lodders}, \&
  {Freedman}}]{2003IAUS..211..345S}
{Saumon}, D., {Marley}, M.~S., {Lodders}, K., \& {Freedman}, R.~S. 2003, in IAU
  Symposium, Vol. 211, Brown Dwarfs, ed. {E.~Mart{\'{\i}}n}, 345

\bibitem[{{Seifahrt} {et~al.}(2009){Seifahrt}, {Helling}, {Burgasser},
  {Allers}, {Cruz}, {Cushing}, {Heiter}, {Looper}, \&
  {Witte}}]{2009AIPC.1094..283S}
{Seifahrt}, A., {et~al.} 2009, in American Institute of Physics Conference
  Series, Vol. 1094, American Institute of Physics Conference Series, ed.
  {E.~Stempels}, 283--290

\bibitem[{{Simcoe} {et~al.}(2008){Simcoe}, {Burgasser}, {Bernstein}, {Bigelow},
  {Fishner}, {Forrest}, {McMurtry}, {Pipher}, {Schechter}, \&
  {Smith}}]{2008SPIE.7014E..27S}
{Simcoe}, R.~A., {et~al.} 2008, in Presented at the Society of Photo-Optical
  Instrumentation Engineers (SPIE) Conference, Vol. 7014, Society of
  Photo-Optical Instrumentation Engineers (SPIE) Conference Series

\bibitem[{{Simcoe} {et~al.}(2010){Simcoe}, {Burgasser}, {Bochanski},
  {Schechter}, {Bernstein}, {Bigelow}, {Pipher}, {Forrest}, {McMurtry},
  {Smith}, \& {Fishner}}]{2010SPIE.7735E..38S}
{Simcoe}, R.~A., {et~al.} 2010, in Society of Photo-Optical Instrumentation
  Engineers (SPIE) Conference Series, Vol. 7735, Society of Photo-Optical
  Instrumentation Engineers (SPIE) Conference Series

\bibitem[{{Skrutskie} {et~al.}(2006){Skrutskie}, {Cutri}, {Stiening},
  {Weinberg}, {Schneider}, {Carpenter}, {Beichman}, {Capps}, {Chester},
  {Elias}, {Huchra}, {Liebert}, {Lonsdale}, {Monet}, {Price}, {Seitzer},
  {Jarrett}, {Kirkpatrick}, {Gizis}, {Howard}, {Evans}, {Fowler}, {Fullmer},
  {Hurt}, {Light}, {Kopan}, {Marsh}, {McCallon}, {Tam}, {Van Dyk}, \&
  {Wheelock}}]{2006AJ....131.1163S}
{Skrutskie}, M.~F., {et~al.} 2006, \aj, 131, 1163

\bibitem[{{Skrutskie} {et~al.}(2010){Skrutskie}, {Jones}, {Hinz}, {Garnavich},
  {Wilson}, {Nelson}, {Solheid}, {Durney}, {Hoffmann}, {Vaitheeswaran},
  {McMahon}, {Leisenring}, \& {Wong}}]{2010SPIE.7735E.118S}
{Skrutskie}, M.~F., {et~al.} 2010, in Society of Photo-Optical Instrumentation
  Engineers (SPIE) Conference Series, Vol. 7735, Society of Photo-Optical
  Instrumentation Engineers (SPIE) Conference Series

\bibitem[{{Stephens} {et~al.}(2009){Stephens}, {Leggett}, {Cushing}, {Marley},
  {Saumon}, {Geballe}, {Golimowski}, {Fan}, \& {Noll}}]{2009ApJ...702..154S}
{Stephens}, D.~C., {et~al.} 2009, \apj, 702, 154

\bibitem[{{Tokunaga} {et~al.}(2002){Tokunaga}, {Simons}, \&
  {Vacca}}]{2002PASP..114..180T}
{Tokunaga}, A.~T., {Simons}, D.~A., \& {Vacca}, W.~D. 2002, \pasp, 114, 180

\bibitem[{{van Dam} {et~al.}(2006){van Dam}, {Bouchez}, {Le Mignant},
  {Johansson}, {Wizinowich}, {Campbell}, {Chin}, {Hartman}, {Lafon}, {Stomski},
  \& {Summers}}]{2006PASP..118..310V}
{van Dam}, M.~A., {et~al.} 2006, \pasp, 118, 310

\bibitem[{{Warren} \& {Brandt}(2008)}]{2008JGRD..11314220W}
{Warren}, S.~G., \& {Brandt}, R.~E. 2008, Journal of Geophysical Research
  (Atmospheres), 113, D14220

\bibitem[{{Warren} {et~al.}(2007){Warren}, {Mortlock}, {Leggett}, {Pinfield},
  {Homeier}, {Dye}, {Jameson}, {Lodieu}, {Lucas}, {Adamson}, {Allard}, {Barrado
  Y Navascu{\'e}s}, {Casali}, {Chiu}, {Hambly}, {Hewett}, {Hirst}, {Irwin},
  {Lawrence}, {Liu}, {Mart{\'{\i}}n}, {Smart}, {Valdivielso}, \&
  {Venemans}}]{2007MNRAS.381.1400W}
{Warren}, S.~J., {et~al.} 2007, \mnras, 381, 1400

\bibitem[{{Weights} {et~al.}(2009){Weights}, {Lucas}, {Roche}, {Pinfield}, \&
  {Riddick}}]{2009MNRAS.392..817W}
{Weights}, D.~J., {Lucas}, P.~W., {Roche}, P.~F., {Pinfield}, D.~J., \&
  {Riddick}, F. 2009, \mnras, 392, 817, erratum

\bibitem[{{Werner} {et~al.}(2004){Werner}, {Roellig}, {Low}, {Rieke}, {Rieke},
  {Hoffmann}, {Young}, {Houck}, {Brandl}, {Fazio}, {Hora}, {Gehrz}, {Helou},
  {Soifer}, {Stauffer}, {Keene}, {Eisenhardt}, {Gallagher}, {Gautier}, {Irace},
  {Lawrence}, {Simmons}, {Van Cleve}, {Jura}, {Wright}, \&
  {Cruikshank}}]{2004ApJS..154....1W}
{Werner}, M.~W., {et~al.} 2004, \apjs, 154, 1

\bibitem[{{Whitworth} {et~al.}(2007){Whitworth}, {Bate}, {Nordlund},
  {Reipurth}, \& {Zinnecker}}]{2007prpl.conf..459W}
{Whitworth}, A., {Bate}, M.~R., {Nordlund}, {\AA}., {Reipurth}, B., \&
  {Zinnecker}, H. 2007, Protostars and Planets V, 459

\bibitem[{{Whitworth} \& {Stamatellos}(2006)}]{2006A&A...458..817W}
{Whitworth}, A.~P., \& {Stamatellos}, D. 2006, \aap, 458, 817

\bibitem[{{Wilson} {et~al.}(2003){Wilson}, {Eikenberry}, {Henderson},
  {Hayward}, {Carson}, {Pirger}, {Barry}, {Brandl}, {Houck}, {Fitzgerald}, \&
  {Stolberg}}]{2003SPIE.4841..451W}
{Wilson}, J.~C., {et~al.} 2003, in Society of Photo-Optical Instrumentation
  Engineers (SPIE) Conference Series, Vol. 4841, Society of Photo-Optical
  Instrumentation Engineers (SPIE) Conference Series, ed. {M.~Iye \&
  A.~F.~M.~Moorwood}, 451--458

\bibitem[{{Wizinowich} {et~al.}(2006){Wizinowich}, {Le Mignant}, {Bouchez},
  {Campbell}, {Chin}, {Contos}, {van Dam}, {Hartman}, {Johansson}, {Lafon},
  {Lewis}, {Stomski}, {Summers}, {Brown}, {Danforth}, {Max}, \&
  {Pennington}}]{2006PASP..118..297W}
{Wizinowich}, P.~L., {et~al.} 2006, \pasp, 118, 297

\bibitem[{{Wright} {et~al.}(2010){Wright}, {Eisenhardt}, {Mainzer}, {Ressler},
  {Cutri}, {Jarrett}, {Kirkpatrick}, {Padgett}, {McMillan}, {Skrutskie},
  {Stanford}, {Cohen}, {Walker}, {Mather}, {Leisawitz}, {Gautier}, {McLean},
  {Benford}, {Lonsdale}, {Blain}, {Mendez}, {Irace}, {Duval}, {Liu}, {Royer},
  {Heinrichsen}, {Howard}, {Shannon}, {Kendall}, {Walsh}, {Larsen}, {Cardon},
  {Schick}, {Schwalm}, {Abid}, {Fabinsky}, {Naes}, \&
  {Tsai}}]{2010AJ....140.1868W}
{Wright}, E.~L., {et~al.} 2010, \aj, 140, 1868

\bibitem[{{York} {et~al.}(2000){York}, {Adelman}, {Anderson}, {Anderson},
  {Annis}, {Bahcall}, {Bakken}, {Barkhouser}, {Bastian}, {Berman}, {Boroski},
  {Bracker}, {Briegel}, {Briggs}, {Brinkmann}, {Brunner}, {Burles}, {Carey},
  {Carr}, {Castander}, {Chen}, {Colestock}, {Connolly}, {Crocker}, {Csabai},
  {Czarapata}, {Davis}, {Doi}, {Dombeck}, {Eisenstein}, {Ellman}, {Elms},
  {Evans}, {Fan}, {Federwitz}, {Fiscelli}, {Friedman}, {Frieman}, {Fukugita},
  {Gillespie}, {Gunn}, {Gurbani}, {de Haas}, {Haldeman}, {Harris}, {Hayes},
  {Heckman}, {Hennessy}, {Hindsley}, {Holm}, {Holmgren}, {Huang}, {Hull},
  {Husby}, {Ichikawa}, {Ichikawa}, {Ivezi{\'c}}, {Kent}, {Kim}, {Kinney},
  {Klaene}, {Kleinman}, {Kleinman}, {Knapp}, {Korienek}, {Kron}, {Kunszt},
  {Lamb}, {Lee}, {Leger}, {Limmongkol}, {Lindenmeyer}, {Long}, {Loomis},
  {Loveday}, {Lucinio}, {Lupton}, {MacKinnon}, {Mannery}, {Mantsch}, {Margon},
  {McGehee}, {McKay}, {Meiksin}, {Merelli}, {Monet}, {Munn}, {Narayanan},
  {Nash}, {Neilsen}, {Neswold}, {Newberg}, {Nichol}, {Nicinski}, {Nonino},
  {Okada}, {Okamura}, {Ostriker}, {Owen}, {Pauls}, {Peoples}, {Peterson},
  {Petravick}, {Pier}, {Pope}, {Pordes}, {Prosapio}, {Rechenmacher}, {Quinn},
  {Richards}, {Richmond}, {Rivetta}, {Rockosi}, {Ruthmansdorfer}, {Sandford},
  {Schlegel}, {Schneider}, {Sekiguchi}, {Sergey}, {Shimasaku}, {Siegmund},
  {Smee}, {Smith}, {Snedden}, {Stone}, {Stoughton}, {Strauss}, {Stubbs},
  {SubbaRao}, {Szalay}, {Szapudi}, {Szokoly}, {Thakar}, {Tremonti}, {Tucker},
  {Uomoto}, {Vanden Berk}, {Vogeley}, {Waddell}, {Wang}, {Watanabe},
  {Weinberg}, {Yanny}, \& {Yasuda}}]{2000AJ....120.1579Y}
{York}, D.~G., {et~al.} 2000, \aj, 120, 1579

\bibitem[{{Yurchenko} {et~al.}(2010){Yurchenko}, {Barber}, \&
  {Tennyson}}]{2010arXiv1011.1569Y}
{Yurchenko}, S.~N., {Barber}, R.~J., \& {Tennyson}, J. 2010, ArXiv e-prints

\bibitem[{{Yurchenko} {et~al.}(2011){Yurchenko}, {Barber}, \&
  {Tennyson}}]{2011MNRAS.413.1828Y}
---. 2011, \mnras, 413, 1828

\end{thebibliography}

\clearpage 

\begin{figure} 
\centerline{\hbox{\includegraphics[width=7in,angle=90]{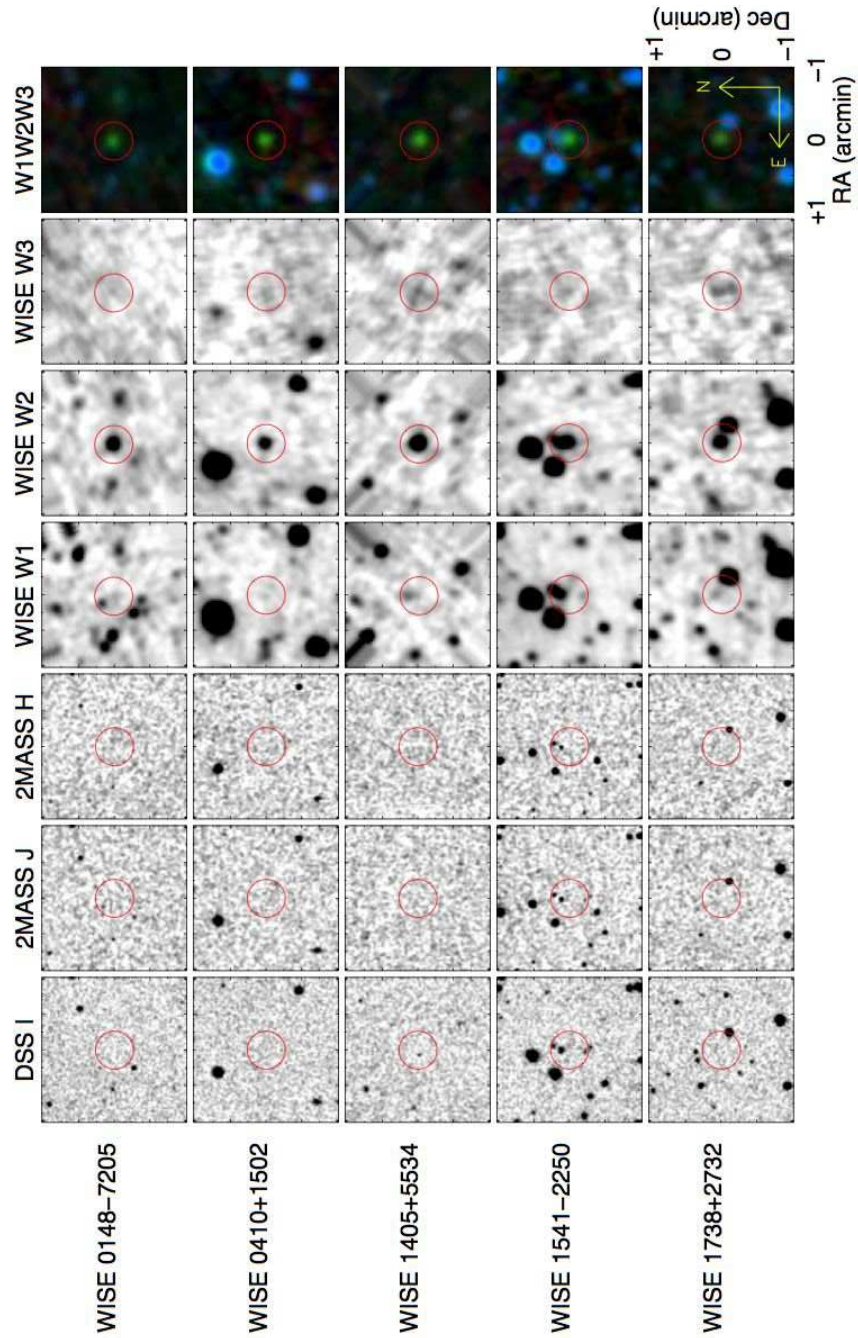}}}
\caption{\label{fig:Finder}2 $\times$ 2 arcmin DSS I, 2MASS $J$ and $H$,
  WISE W1, W2, and W3, and a W1W2W3 false color composite of the five
  new WISE brown dwarfs.  In the color composite images on the far
  right, the W1, W2, and W3 are bands are color coded blue, green, and
  red, respectively.}
\end{figure}

\clearpage

\begin{figure} 
\centerline{\hbox{\includegraphics[width=7in,angle=90]{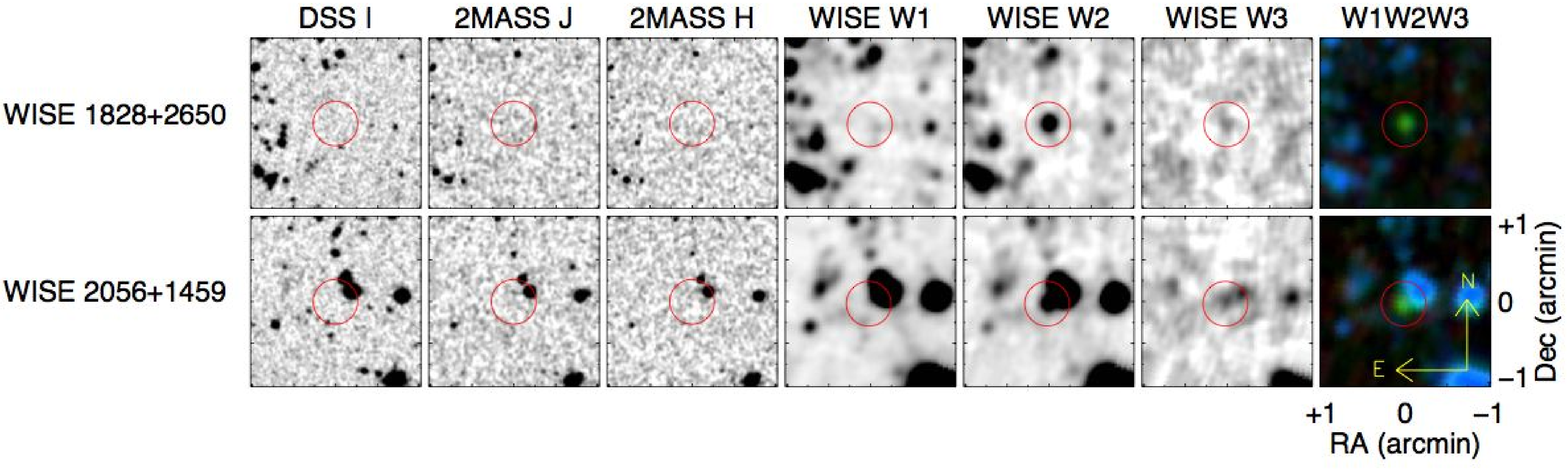}}}
\end{figure}

\clearpage

\begin{figure} 
\centerline{\hbox{\includegraphics[width=6in,angle=0]{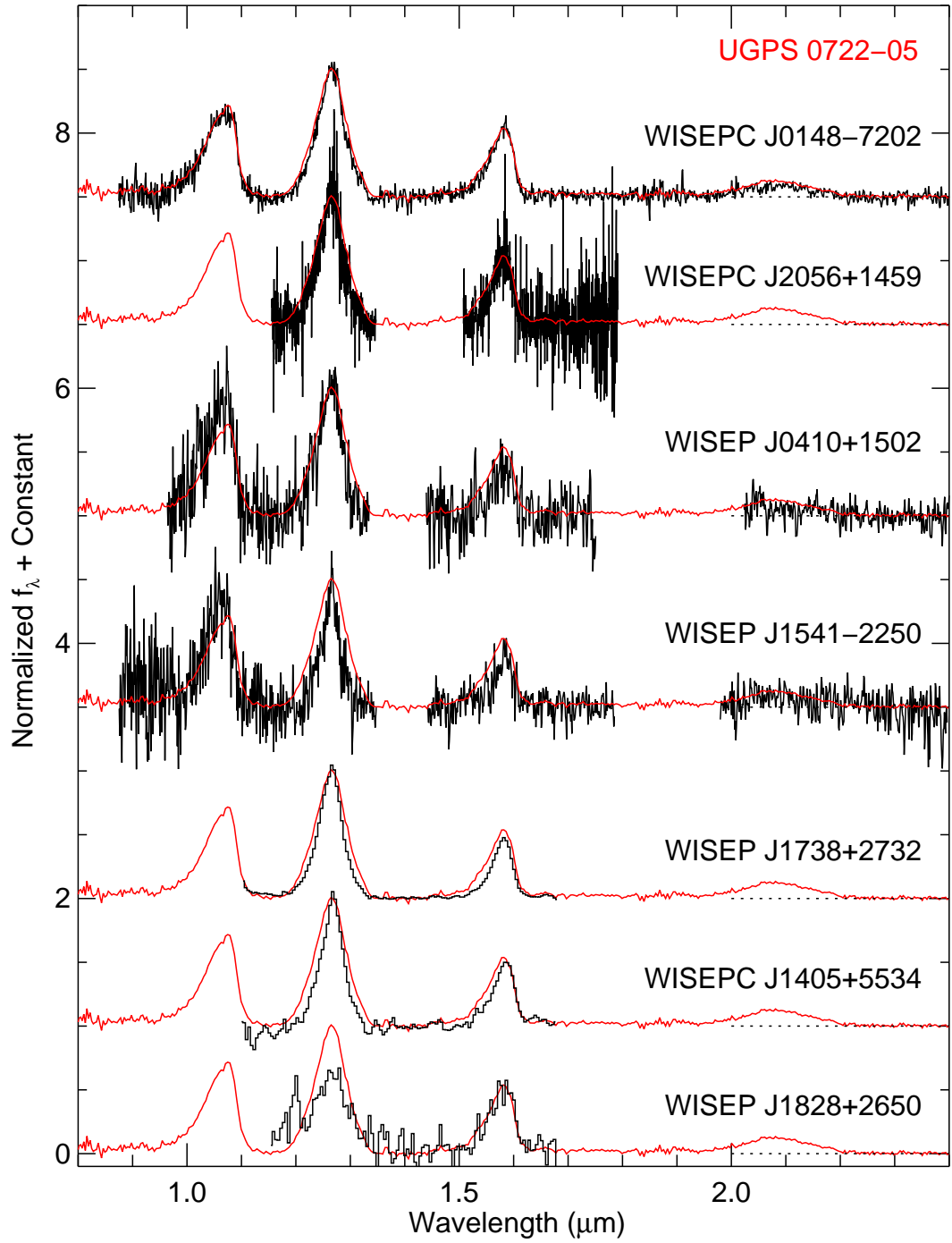}}}
\caption{\label{fig:rawseq}Near-infrared spectra of the new WISE brown
  dwarfs (black) as compared to the spectrum of \UGPSzeroseventwotwo\
  (red).  The data have been normalized to unity at the peak of the $J$
  band (except for \Weighteentwentyeight\ which is normalized to unity
  at the peak of the $H$-band) and offset by constants (dotted lines).}
\end{figure}

\clearpage 

\begin{figure} 
\centerline{\hbox{\includegraphics[width=6in,angle=0]{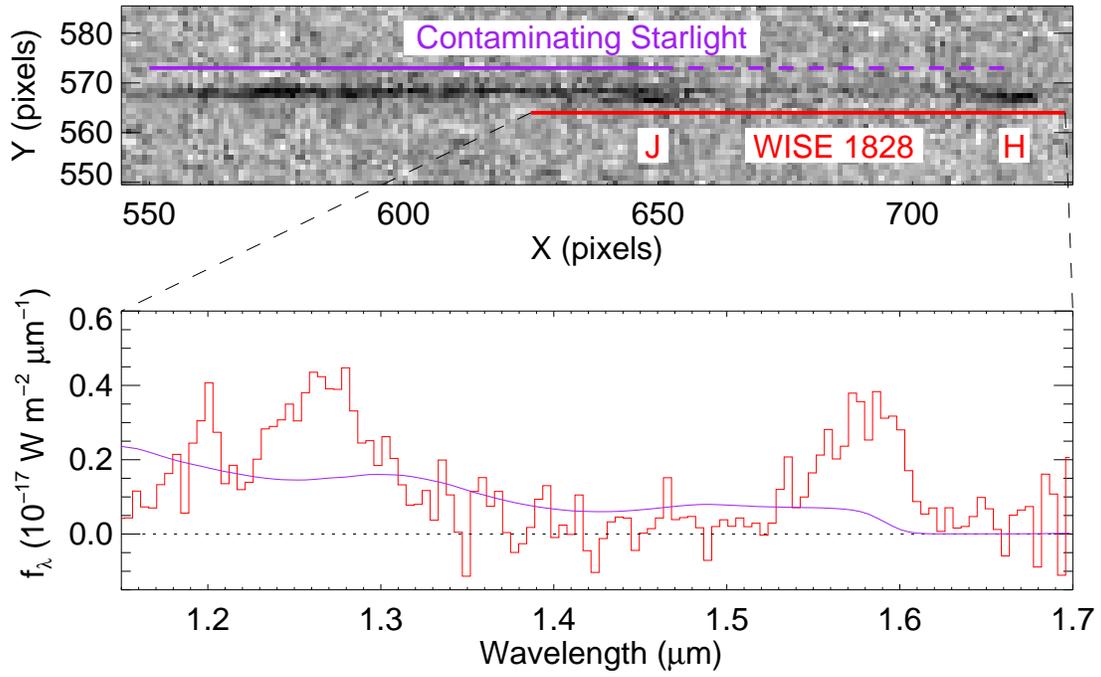}}}
\vspace{0.3in}
\caption{\label{fig:1828img}Top: Subimage of the drizzled WFC3/HST grism
  image centered on the position of \Weighteentwentyeight.  The location
  of the spectrum of \Weighteentwentyeight\ is indicated in red along
  with the positions of the $J$- and $H$-band peaks.  The location of
  the contaminating starlight is shown in purple and consists of second
  and third order light from two other stars in the WFC3 field of view.
  Bottom: The spectrum of \Weighteentwentyeight\ (red) and the
  contamination spectrum (purple).  The stellar contamination becomes
  progressively worse at shorter wavelengths.}
\end{figure}

\clearpage

\begin{figure} 
\centerline{\hbox{\includegraphics[width=6in,angle=0]{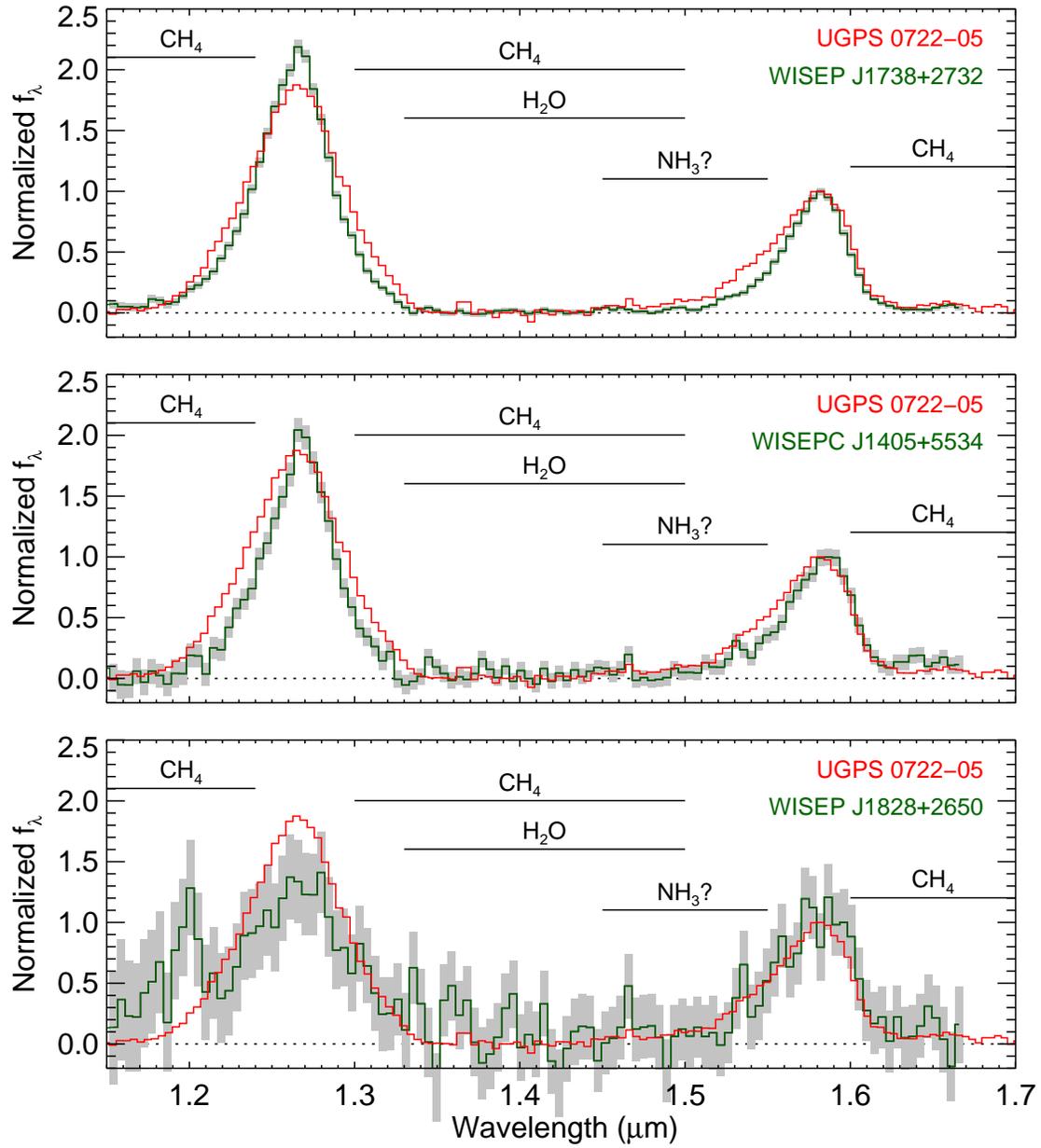}}}
\caption{\label{fig:t9y0comp}1.15$-$1.70 $\mu$m spectra of
  \Wseventeenthirtyeight, \Wfourteenzerofive, and \Weighteentwentyeight\
  along with the spectrum \UGPSzeroseventwotwo.  The uncertainties in
  the spectra are shown as gray bars. The spectra were all normalized to
  unity at the peak of the $H$-band (1.58 $\mu$m).  Prominent molecular
  absorption bands are indicated.}
\end{figure}

\clearpage 

\begin{figure} 
\centerline{\hbox{\includegraphics[width=6in,angle=0]{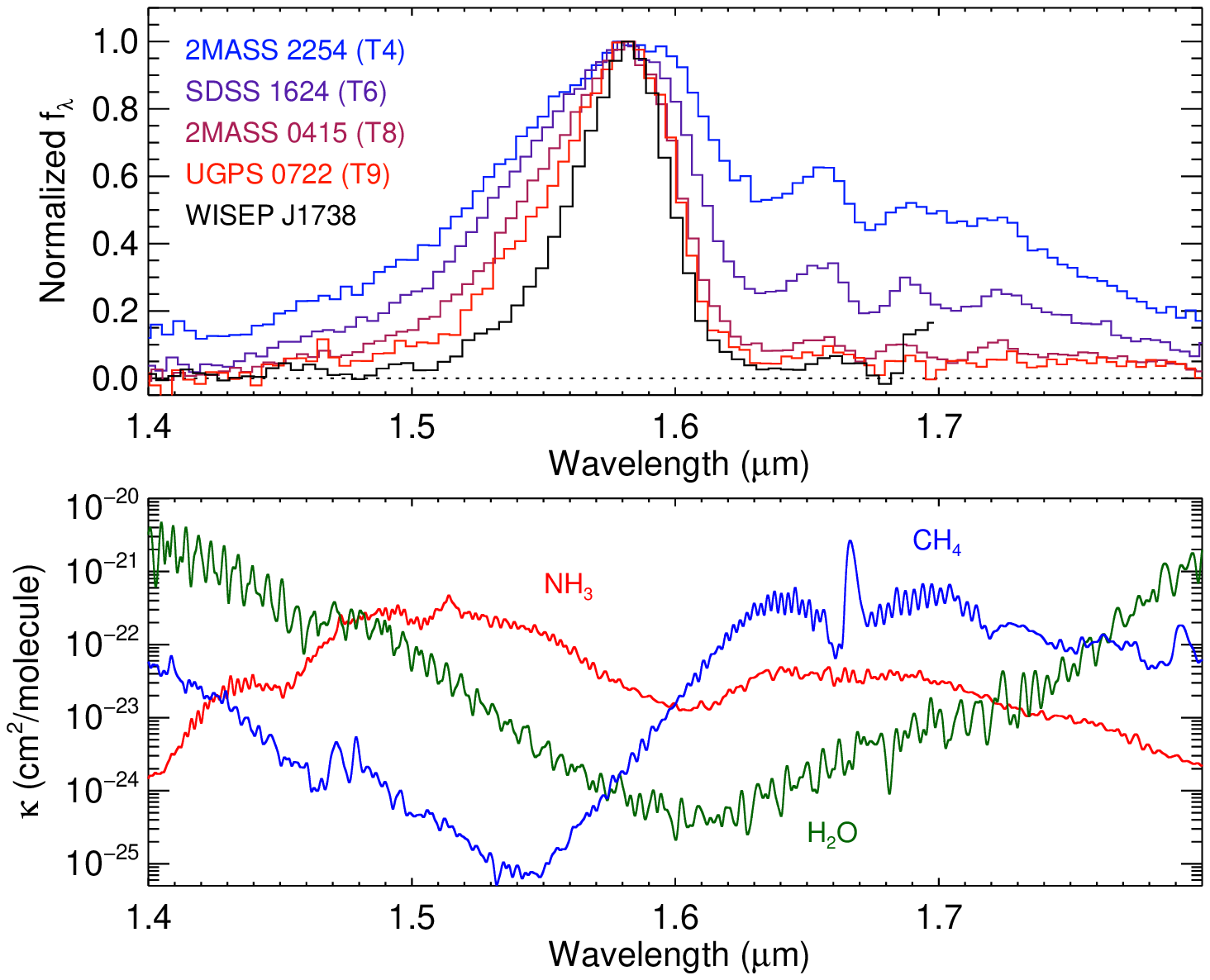}}}
\caption{\label{fig:nh3comp}Top: $H$-band spectrum of \TfourFull,
  \TsixFull\, and \TeightFull, the T4, T6, and T8 spectral standards
  \citep{2006ApJ...637.1067B}, along with the spectrum of
  \UGPSzeroseventwotwo, and \Wseventeenthirtyeight.  The spectra have
  been normalized to unity at their peak flux values.  Bottom: Opacity
  data computed in chemical equilibrium for NH$_3$
  \citep{2010arXiv1011.1569Y}, H$_2$O \citep{2008ApJS..174..504F}, and
  CH$_4$ \citep{2008ApJS..174..504F} at $T$=600 K and $P$=1 bar.  Note
  that the change in the spectral morphology of the blue wing of the
  $H$-band peak is similar between T6/T8 and T8/T9 suggesting a common
  absorber or set of absorbers.  In contrast, the spectrum of
  \Wseventeenthirtyeight\ exhibits excess absorption that matches the
  position of the NH$_3$ absorption shown in the lower panel.}
\end{figure}

\clearpage

\begin{figure} 
\centerline{\hbox{\includegraphics[width=6in,angle=0]{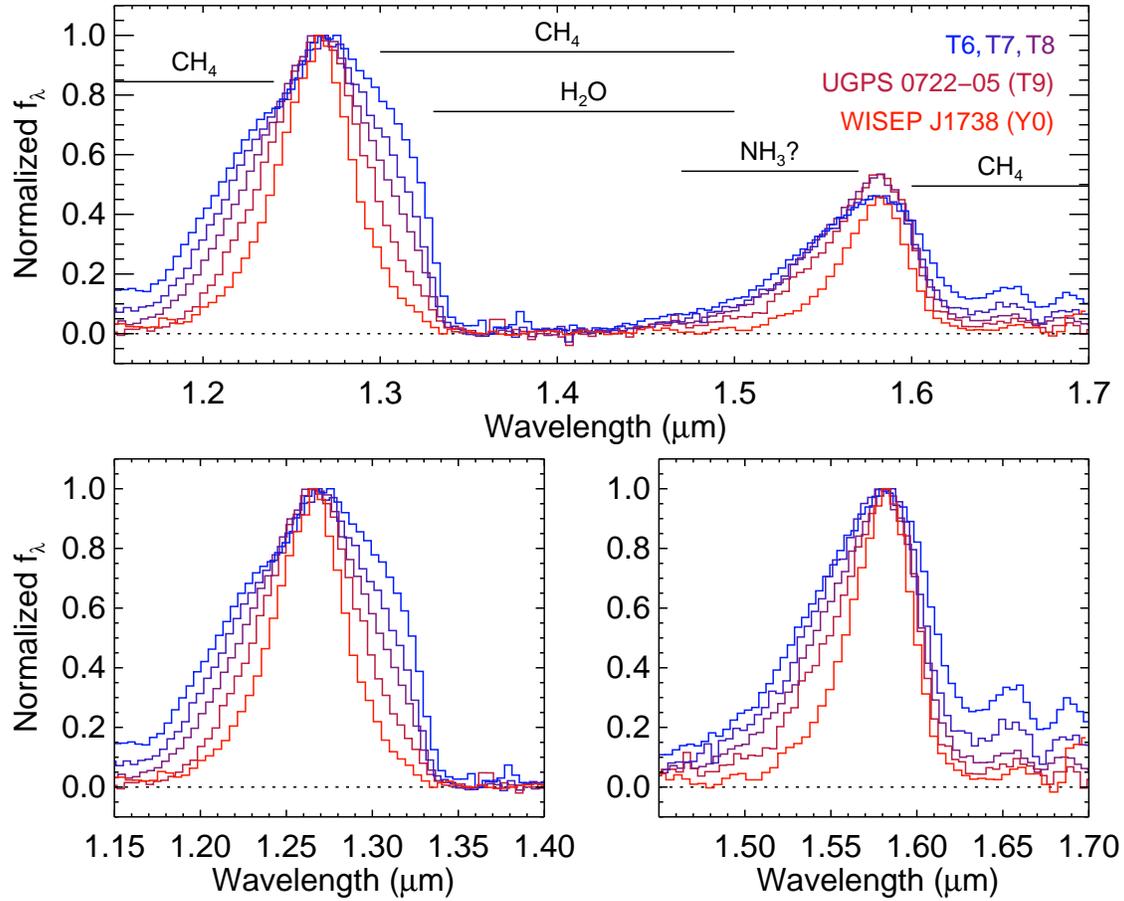}}}
\caption{\label{fig:t6t7t8t9y0}IRTF/SpeX spectra of the
  \citet{2006ApJ...637.1067B} spectral standards, \TsixFull\ (T6),
  \TsevenFull\ (T7), and \TeightFull\ (T8), our IRTF/SpeX spectrum of
  \UGPSzeroseventwotwo\, and the WFC3/\textit{HST} spectrum of
  \Wseventeenthirtyeight.  The spectra have been normalized to unity at
  their peak flux level in each panel.  Prominent molecular absorption
  bands are indicated in the top panel.}
\end{figure}

\clearpage

\begin{figure} 
\centerline{\hbox{\includegraphics[width=6in,angle=0]{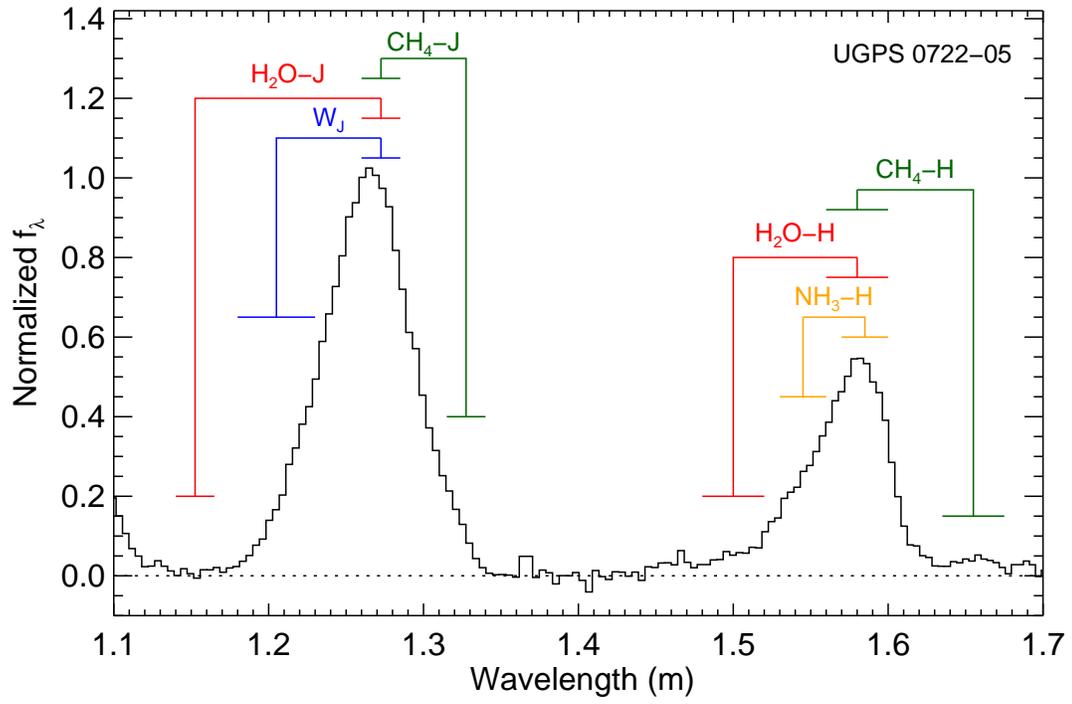}}}
\caption{\label{fig:idxkey}Illustration of the H$_2$O-$J$, W$_J$,
  CH$_4$-$J$, CH$_4$-$J$, NH$_3$-$H$, and CH$_4$-$H$ indices overplotted
  on the spectrum UGPS 0722-05.}
\end{figure}

\clearpage

\begin{figure} 
  \centerline{\hbox{\includegraphics[width=6in,angle=0]{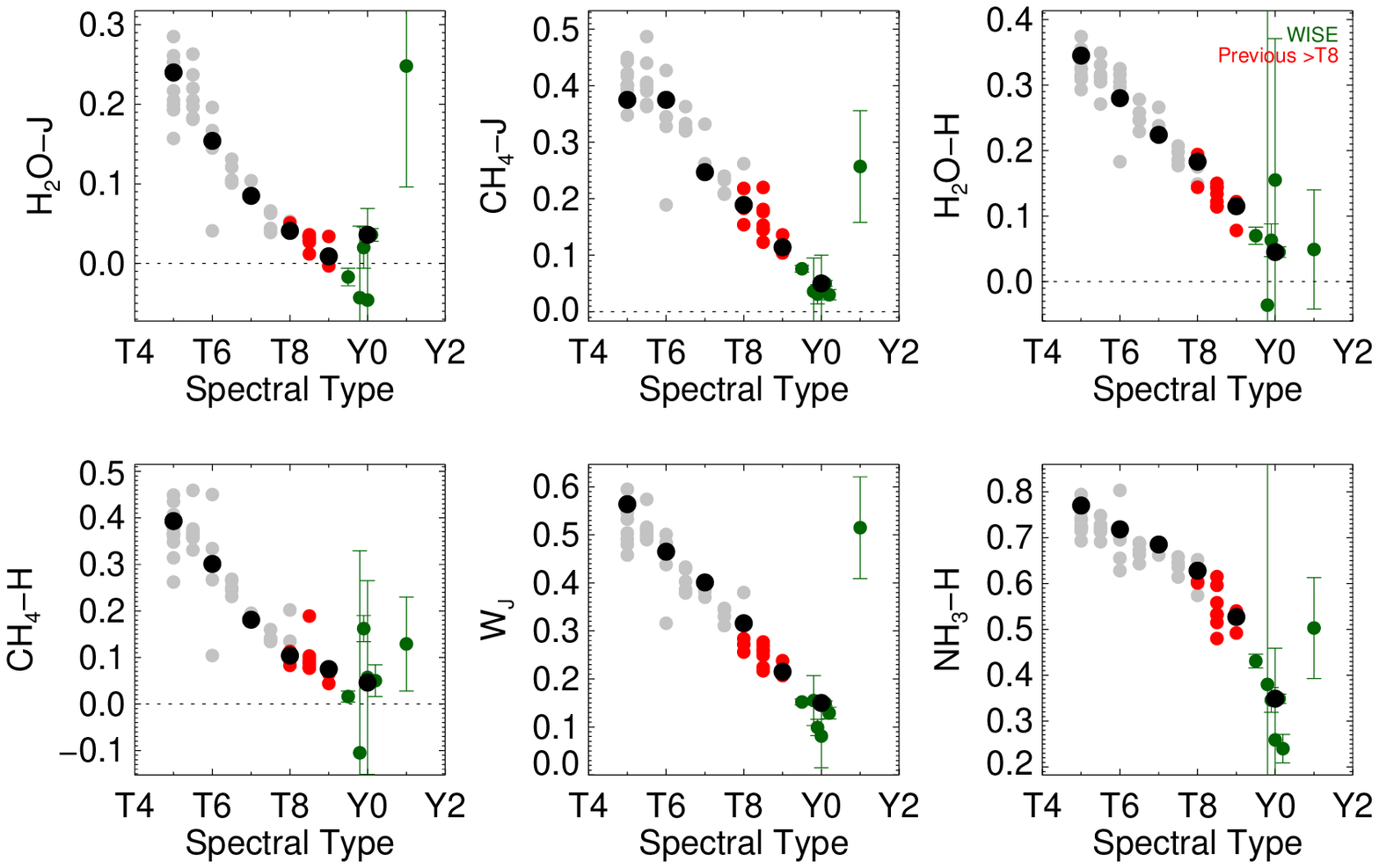}}}
  \caption{\label{fig:indices}Values of the H$_2$O$-J$, CH$_4$$-$$J$,
    H$_2$O$-$$H$, CH$_4$$-$$H$ \citep{2006ApJ...637.1067B}, $W_J$
    \citep{2007MNRAS.381.1400W}, and NH$_3$$-$$H$
    \citep{2008A&A...482..961D} spectral indices as a function of
    spectral type.  The black points are for the T6$-$Y0 spectral
    standards.  The grey points were computed using spectra of late-type
    (T5$-$T8) T dwarfs from the SpeX Prism Spectral Library.  The red
    points are the twelve T dwarfs with previously published spectral
    types later than T8 and the green points are the remaining six WISE
    brown dwarfs.  For plotting purposes only, we have assigned
    \Weighteentwentyeight\ a spectral type of Y1.}
\end{figure}

\clearpage

\begin{figure} 
  \centerline{\hbox{\includegraphics[width=6in,angle=0]{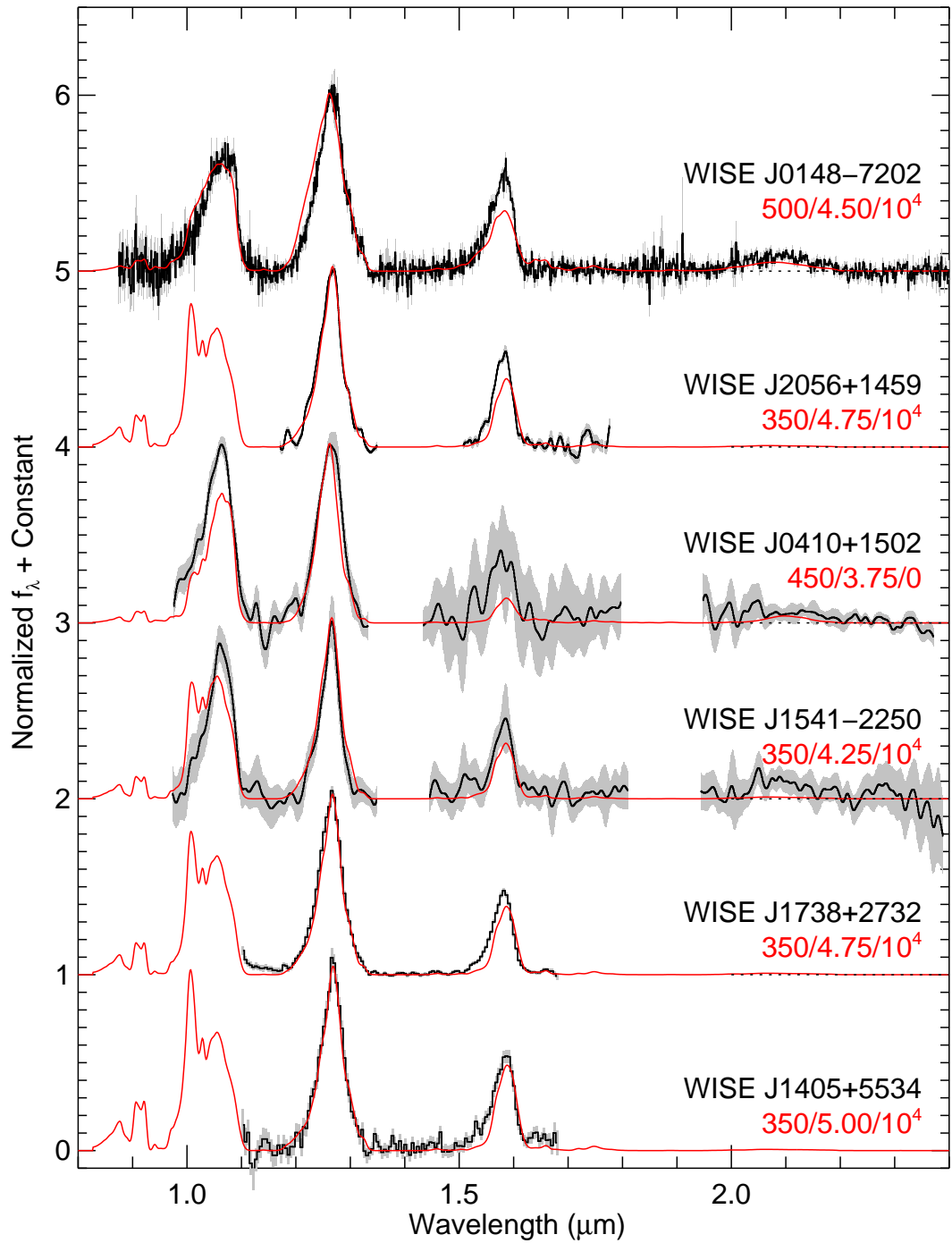}}}
  \caption{\label{fig:modcomp}Best fitting models (red) overplotted on
    the near-infrared spectra of six of the seven new WISE brown dwarfs
    (black).  The spectra were normalized to unity at the peak flux in
    the $J$-band and offset by constants (dotted lines).  The
    uncertainties in the spectra are given by grey bars.  The best
    fitting model parameters are given in the form \teff (K)/\logg\ (cm
    s$^{-2}$)/\kzz\ (cm$^2$ s$^{-1}$).}
\end{figure}

\clearpage

\begin{figure} 
  \centerline{\hbox{\includegraphics[width=6in,angle=0]{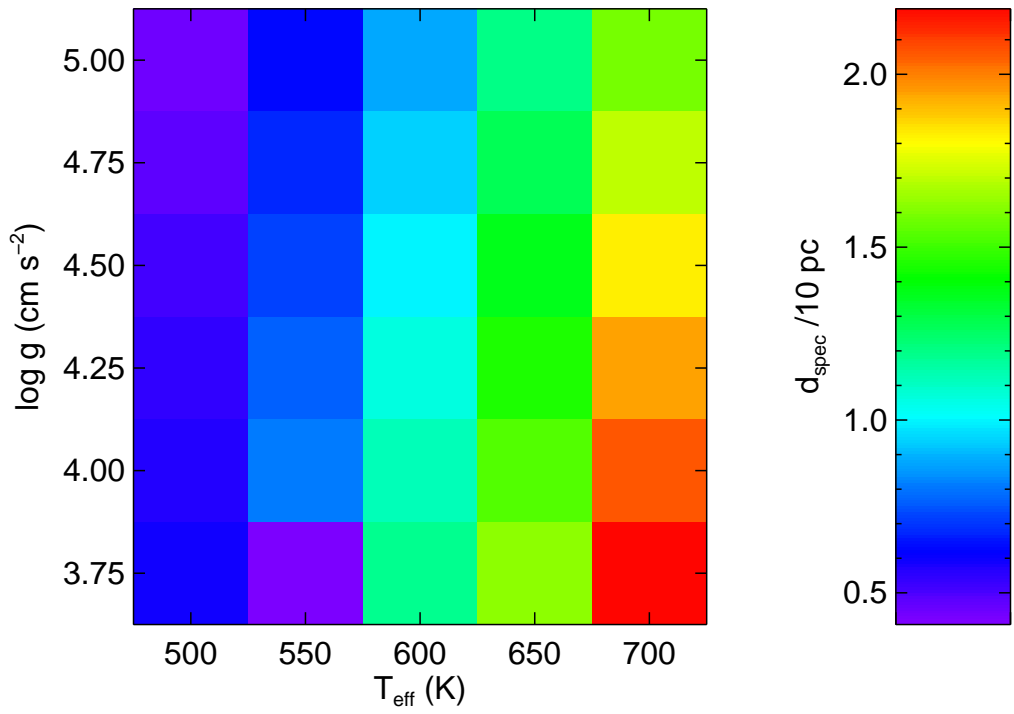}}}
  \caption{\label{fig:distcomp}Impact of systematic errors in the
    derived values of (\teff, \logg) on the spectroscopic distance,
    $d_\mathrm{spec}$, for a hypothetical dwarf with \teff=600 K and
    \logg = 4.5 cm s$^{-2}$.}
\end{figure}

\begin{figure} 
  \centerline{\hbox{\includegraphics[width=3.5in,angle=0]{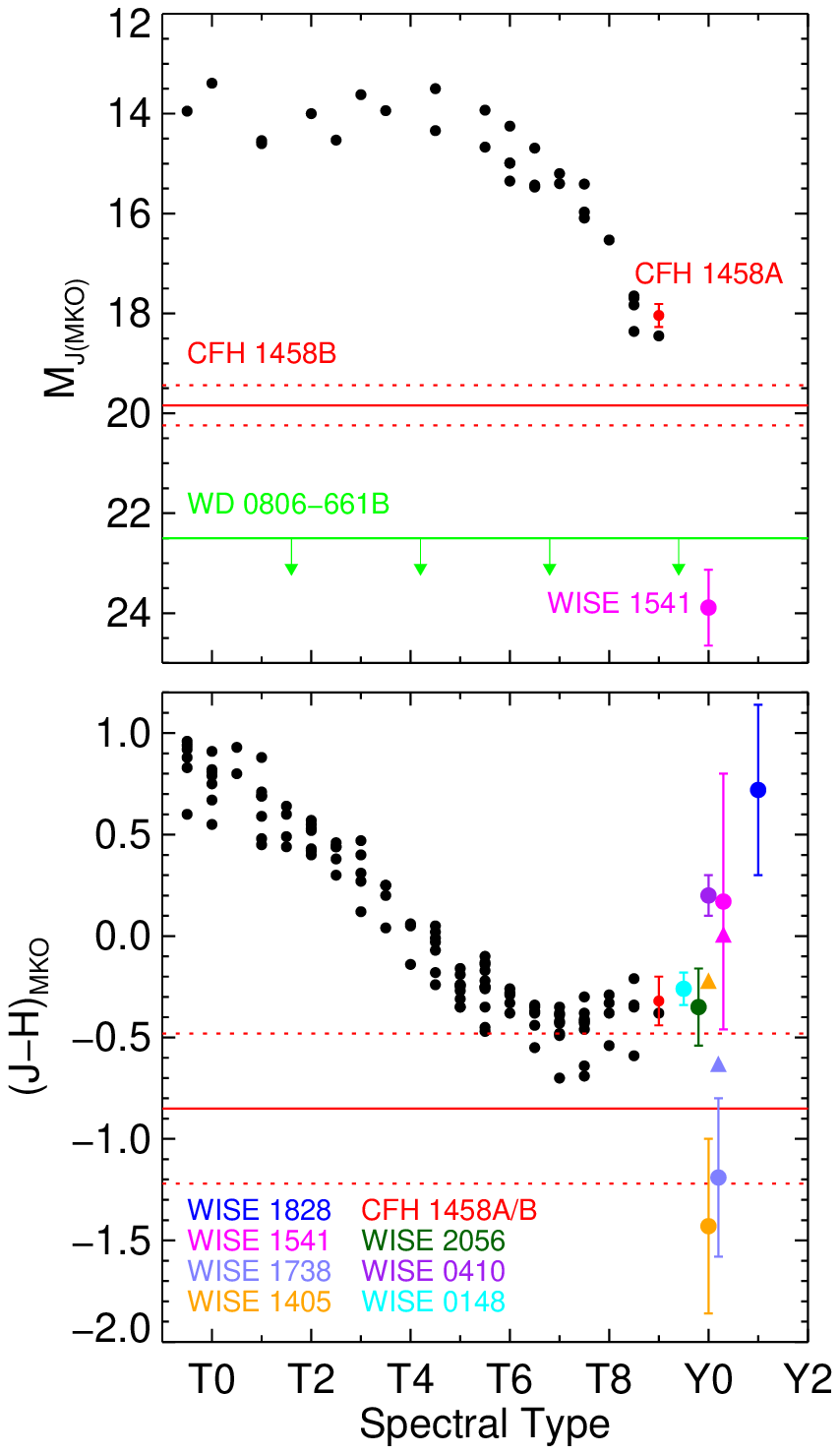}}}
  \caption{\label{fig:bdcomp}Comparison of the absolute $J$-band
    magnitudes and $J-H$ colors of the WISE brown dwarfs,
    \WDzeroeightzerosixFull B, and \CFHfourteenfiftyeight B.  The field
    population (black circles) is from the compilation of
    \citet{2010ApJ...710.1627L} and the spectral types of the late-type
    T dwarfs have been changed to match the subtypes given in Table
    \ref{tab:gtt8}.  Synthetic colors computed by integrating the $J$
    and $H$ bandpasses over spectra are plotted as triangles.  For
    plotting purposes only, we have assigned a spectral type of Y1 for
    \Weighteentwentyeight.}
\end{figure}

\clearpage

%
%=======================================  WISE Photometry ==============================================
%

\begin{deluxetable}{rlllll}
\tablecolumns{6}
\tabletypesize{\scriptsize} 
\tablewidth{0pc}
\tablecaption{\label{tab:wisephot}WISE Photometry}
\tablehead{
\colhead{Object} & 
\colhead{W1} & 
\colhead{W2} & 
\colhead{W3} & 
\colhead{W4} & 
\colhead{W1$-$W2} \\
\colhead{} & 
\colhead{(mag)} & 
\colhead{(mag)} & 
\colhead{(mag)} & 
\colhead{(mag)} & 
\colhead{(mag)}}

\startdata

\WzeroonefoureightFull\       & 18.812$\pm$0.529\tablenotemark{a} & 14.584$\pm$0.052 & $>$12.579                         & $>$9.521                         & 4.228$\pm$0.532 \\
\WzerofourtenFull\            & $>$18.101                         & 14.190$\pm$0.059 & 12.472$\pm$0.482\tablenotemark{a} & $>$8.923                         & $>$3.911$\pm$0.059 \\
\WfourteenzerofiveFull\       & $>$17.989                         & 14.085$\pm$0.041 & 12.312$\pm$0.252                  & $>$9.115                         & $>$3.904$\pm$0.041 \\
\WfifteenfortyoneFull\        & $>$17.018                         & 13.982$\pm$0.112 & 12.134$\pm$0.443\tablenotemark{a} & $>$9.064                         & $>$3.036$\pm$0.112 \\
\WseventeenthirtyeightFull\   & 18.155$\pm$0.362                  & 14.535$\pm$0.057 & 12.536$\pm$0.350                  & $>$9.182                         & 3.620$\pm$0.366 \\
\WeighteentwentyeightFull\    & $>$18.452                         & 14.276$\pm$0.050 & 12.320$\pm$0.291                  & 9.147$\pm$0.438\tablenotemark{a} & $>$4.176$\pm$0.050 \\
\WtwentyfiftysixFull\         & $>$17.742                         & 13.852$\pm$0.043 & 11.791$\pm$0.222                  & $>$8.646                         & $>$3.890$\pm$0.043 \\

\enddata
\tablecomments{Objects designated as WISEP are from the Preliminary
  Release Source Catalog while objects designated as WISEPC are from the
  first-pass processing operations coadd Source Working Database.
  Magnitudes are in the Vega system and are based on profile fits
  (w1mpro, w2mpro, w3mpro, w4mpro).  Upper limits are at the 95\%
  confidence level (see
  \url{http://wise2.ipac.caltech.edu/docs/release/prelim/expsup/sec4\_3c.html\#ul2}).}
\tablenotetext{a}{S/N$\leq$3}

\end{deluxetable}

%
%======================================= NIR Photometry ====================================================
%

\begin{deluxetable}{lcccccccc}
\tablecolumns{9}
\tabletypesize{\scriptsize} 
\tablewidth{0pc}
\tablecaption{\label{tab:nirphot}Near-Infrared Photometry}
\tablehead{
\colhead{Object} & 
\colhead{Filter} & 
\colhead{Instrument} & 
\colhead{Magnitude} & 
\colhead{Exposure} & 
\colhead{Coadds} & 
\colhead{Number} & 
\colhead{Total} & 
\colhead{Date} \\
\colhead{} & 
\colhead{} & 
\colhead{} & 
\colhead{(mag)} & 
\colhead{Time (sec)} &
\colhead{} & 
\colhead{of Images} & 
\colhead{Exp. (sec)} & 
\colhead{(UT)}}

\startdata

\Wzeroonefoureight\       &  $J$  &  PANIC     &  18.96$\pm$0.07  &  30   &  1   &  18   &  540    &  2010 Aug 01  \\
                          &  $H$  &  PANIC     &  19.22$\pm$0.04  &  15   &  1   &  108  &  1620   &  2010 Aug 01  \\
\Wzerofourten\            &  $J$  &  WIRC      &  19.25$\pm$0.05  &  60   &  1   &  15   &  900    &  2010 Aug 29  \\
                          &  $H$  &  WIRC      &  19.05$\pm$0.09  &  30   &  4   &  15   &  1800   &  2010 Jul 26  \\
\Wfourteenzerofive\       &  $J$  &  WIRC      &  20.20$\pm$0.13  &  30   &  2   &  15   &  1800   &  2010 Jul 26  \\
                          &  $H$  &  WIRC      &  21.45$\pm$0.41  &  $\cdots$  & $\cdots$     &  $\cdots$     &  5400   &  multiple     \\
\Wfifteenfortyone\        &  $J$  &  NEWFIRM   &  21.16$\pm$0.36  &  30   &  2   &  10   &  600    &  2011 Apr 17  \\ 
                          &  $H$  &  NEWFIRM   &  20.99$\pm$0.52  &  5    &  12  &  10   &  600    &  2011 Apr 17  \\
\Wseventeenthirtyeight\   &  $J$  &  WIRC      &  19.47$\pm$0.08  &  60   &  1   &  15   &  900    &  2010 Jul 26  \\
                          &  $H$  &  WIRC      &  20.66$\pm$0.38  &  30   &  2   &  15   &  900    &  2010 Jul 26  \\
\Weighteentwentyeight\    &  $J$  &  NIRC2     &  23.57$\pm$0.35  &  120  &  1   &  6    &  720    & 2010 Jul 01   \\
                          &  $H$  &  NIRC2     &  22.85$\pm$0.24  &  120  &  1   &  9    &  1080   & 2010 Jul 01   \\
\Wtwentyfiftysix\         &  $J$  &  WIRC      &  19.31$\pm$0.12  &  60   &  1   &  15   &  900    & 2010 Aug 29   \\
                          &  $H$  &  WIRC      &  $>$19.5         &  30   &  2   &  15   &  900    & 2010 Aug 29   \\
                          &  $J$  &  NIRC2     &  19.21$\pm$0.07  &  120  &  1   &  9    & 1080    & 2010 Jul 01   \\
                          &  $H$  &  NIRC2     &  19.56$\pm$0.18  &  120  &  1   &  6    &  720    & 2010 Jul 01   \\

%&  $J$  &  FIRE      &  20.85$\pm$0.18  &  45   &  1   &  6    &   270   &  2011 Mar 27  \\
\enddata

\end{deluxetable}

%
%===================================================== Spectroscopy Log =============================================================
%

\begin{deluxetable}{rlclccc}
\tablecolumns{7}
\tabletypesize{\scriptsize} 
\tablewidth{0pc}
\tablecaption{\label{tab:speclog}Spectroscopy Log}
\tablehead{
\colhead{Object} & 
\colhead{Instrument} & 
\colhead{UT Date} & 
\colhead{Mode} & 
\colhead{Slit Width} & 
\colhead{Int. Time} & 
\colhead{A0 V Calibrator} \\
\colhead{} & 
\colhead{} & 
\colhead{} & 
\colhead{} & 
\colhead{(arcsec)} & 
\colhead{(sec)} & 
\colhead{Star}}

\startdata

\Wzeroonefoureight\       &  FIRE/Magellan    &  2010 Sep 18    &  Longslit    & 0.6      & 960   & HD 1881    \\
\Wzerofourten\            & FIRE/Magellan     &  2010 Nov 18    &  Longslit    & 1.0      & 600   & HD 18620   \\
\UGPSzeroseventwotwo\     & SpeX/IRTF         &  2011 Jan 26    &  LowRes15    & 0.5      & 1440  & HD 50931   \\
\Wfourteenzerofive\       & WFC3/\textit{HST} &  2011 Mar 14    &  G141        & $\cdots$ & 2212  & $\cdots$   \\
\Wfifteenfortyone\        & FIRE/Magellan     &  2011 Mar 27    &  Longslit    & 0.6      & 1522  & HD 130755  \\
\Wseventeenthirtyeight\   & WFC3/\textit{HST} &  2011 May 12    &  G141        & $\cdots$ & 2012  & $\cdots$   \\
\Weighteentwentyeight\    & WFC3/\textit{HST} &  2011 May 09    &  G141        & $\cdots$ & 2012  & $\cdots$   \\
\Wtwentyfiftysix\         & NIRSPEC/Keck      &  2010 Oct 21    &  low-res (N3)          & 0.38     & 2400  & HD 198070  \\
                          & NIRSPEC/Keck      &  2010 Nov 22    &  low-res (N5)          & 0.38     & 1800  & HD 198069 \\

\enddata
\end{deluxetable}

%
%=======================================  Spectral Indices ==============================================
%

\begin{deluxetable}{rllcccccc}
%\rotate
\tablecolumns{9}
\tabletypesize{\tiny} 
\tablewidth{0pc}
\tablecaption{\label{tab:spectralindices}Spectral Indices}
\tablehead{
\colhead{Object} & 
\colhead{Spectral} & 
\colhead{H$_2$O$-J$} & 
\colhead{CH$_4$$-J$} & 
\colhead{H$_2$O$-H$} & 
\colhead{CH$_4$$-H$} & 
\colhead{W$_J$} & 
\colhead{NH$_3$$-H$} \\
\colhead{} & 
\colhead{Type} & 
\colhead{} & 
\colhead{} & 
\colhead{} & 
\colhead{} & 
\colhead{} & 
\colhead{}}

\startdata

\UGPSzeroseventwotwo\tablenotemark{a}     & T9        & $+$0.009 (0.004) & $+$0.115 (0.003) & $+$0.115 (0.007) & $+$0.075 (0.005) & $+$0.215 (0.003) & $+$0.527 (0.008) \\
\Wzeroonefoureight\                       & T9.5      & $-$0.017 (0.011) & $+$0.076 (0.006) & $+$0.070 (0.013) & $+$0.016 (0.012) & $+$0.152 (0.006) & $+$0.431 (0.015) \\
\Wzerofourten\                            & Y0        & $-$0.043 (0.090) & $+$0.036 (0.059) & $-$0.036 (0.458) & $-$0.105 (0.434) & $+$0.155 (0.052) & $+$0.380 (0.574) \\
\Wfourteenzerofive\                       & Y0 (pec?) & $+$0.020 (0.026) & $+$0.031 (0.017) & $+$0.063 (0.025) & $+$0.162 (0.028) & $+$0.099 (0.017) & $+$0.346 (0.027) \\
\Wfifteenfortyone\                        & Y0        & $-$0.046 (0.115) & $+$0.040 (0.060) & $+$0.155 (0.216) & $+$0.057 (0.208) & $+$0.081 (0.066) & $+$0.259 (0.200) \\
\Wseventeenthirtyeight\                   & Y0        & $+$0.036 (0.008) & $+$0.050 (0.005) & $+$0.045 (0.008) & $+$0.050 (0.009) & $+$0.149 (0.005) & $+$0.349 (0.010) \\
\Weighteentwentyeight\                    & $>$Y0     & $+$0.248 (0.152) & $+$0.257 (0.099) & $+$0.049 (0.091) & $+$0.129 (0.101) & $+$0.515 (0.106) & $+$0.503 (0.110) \\
\Wtwentyfiftysix\                         & Y0        & $\cdots$         & $+$0.030 (0.009) & $\cdots$         & $+$0.050 (0.034) & $+$0.129 (0.012) & $+$0.240 (0.031) \\
\enddata
\tablecomments{The H$_2$O$-J$ and H$_2$O$-H$ indices cannot be computed
  for \WtwentyfiftysixFull\ because its spectrum does not span the
  entire wavelength range of the indices.}

\tablenotetext{a}{The values differ from that measured by
  \citet{2010MNRAS.408L..56L}.  Our two spectra agree well except deep
  in the CH$_4$ and H$_2$O absorption bands, where our spectrum exhibits
  lower flux levels.  The reason for this discrepancy is unclear but it
  may be a result of the fact that the \citeauthor{2010MNRAS.408L..56L}
  spectrum was created by merging separate spectra that were absolutely
  flux calibrated using near-infrared photometry.}.

\end{deluxetable}

%
%===================================================  gt T8 Log =====================================================================
%

\begin{deluxetable}{lcll}
\tablecolumns{10}
\tabletypesize{\scriptsize} 
\tablewidth{0pc}
\tablecaption{\label{tab:gtt8}Previously Published Brown Dwarfs with Spectral Types Later than T8}
\tablehead{
\colhead{Object} & 
\colhead{Previous} & 
\colhead{Reference} & 
\colhead{Adopted} \\
\colhead{} & 
\colhead{Spectral Type} &
\colhead{} & 
\colhead{Spectral Type}}

\startdata

Ross 458C                      & T8          & \citet{2010ApJ...725.1405B}  & T8 \\
                               & T8.5p       & \citet{2011MNRAS.414.3590B}  & $\cdots$ \\
\ULAStwelvethirtyeightFull\    & T8.5        & \citet{2008MNRAS.391..320B}  & T8 \\
ULAS J130217.21$+$130851.2     & T8.5        & \citet{2010MNRAS.406.1885B}  & T8 \\
\ULASzerozerothirtyfourFull\   & T8.5        & \citet{2007MNRAS.381.1400W}  & T8.5 \\
                               & T9          & \citet{2008MNRAS.391..320B}  & $\cdots$ \\
\CFHBDSzerozerofivenineFull\   & $\gtrsim$T9 & Delorme et al. (2008)        & T8.5 \\
                               & T9          & \citet{2008MNRAS.391..320B}  & $\cdots$ \\
\WzerofourfiveeightFull\       & T9          & \citet{2011ApJ...726...30M}  & T8.5 \\
UGPS J052127.27$+$364048.6     & T8.5        & \citet{2011MNRAS.414L..90B}  & T8.5 \\
\ULASthirteenthirtyfiveFull\   & T9          & \citet{2008MNRAS.391..320B}  & T8.5 \\
Wolf 940B                      & T8.5        & \citet{2009MNRAS.395.1237B}  & T8.5 \\
WISEPC J181210.85$+$272144.3   & T8.5:       & \citet{2011ApJ...735..116B}  & T8.5:    \\
\UGPSzeroseventwotwoFull\      & T10         & \citet{2010MNRAS.408L..56L}  & T9  \\
\CFHfourteenfiftyeightFull AB  & T9.5        & Liu et al. (2011)            & T9 \\

\enddata

\end{deluxetable}

%
%=========================================  Model Parameters ========================================================
%
\begin{deluxetable}{rlllcll}
\tablecolumns{8}
\tabletypesize{\scriptsize} 
\tablewidth{0pc}
\tablecaption{\label{tab:properties}Atmospheric and Structural Properties}
\tablehead{
\colhead{Object} & 
\colhead{SpType} & 
\colhead{$T_\mathrm{eff}$} & 
\colhead{$\log$ $g$} & 
\colhead{log $K_\mathrm{zz}$} & 
\colhead{$R$} & 
\colhead{$M$} \\
\colhead{} & 
\colhead{} & 
\colhead{(K)} & 
\colhead{(cm s$^{-2}$)} & 
\colhead{(cm$^2$ s$^{-1}$)} & 
\colhead{($R_\mathrm{Jup}$)} & 
\colhead{($M_\mathrm{Jup}$)}}

\startdata
\UGPSzeroseventwotwo\     & T9         & 650                       & 4.00 (4.00$-$4.25)  & 4       & 1.21 (1.14$-$1.21) & 6 (6$-$9)     \\
\Wzeroonefoureight\       & T9.5       & 500 (500$-$500)           & 4.50 (4.50$-$4.75)  & 4       & 1.04 (0.96$-$1.04) & 13 (13$-$21)  \\ 
\Wzerofourten\            & Y0         & 450 (400$-$500)           & 3.75 (3.75$-$4.25)  & 0       & 1.22 (1.09$-$1.22) & 3 (3$-$9)     \\ 
\Wfourteenzerofive\       & Y0 (pec?)  & 350                       & 5.00                & 4       & 0.86               & 30                \\ 
\Wfifteenfortyone\        & Y0         & 350                       & 4.50 (4.25$-$4.5)   & 4       & 1.01 (1.01$-$1.07) & 12 (8$-$12)   \\ 
\Wseventeenthirtyeight\   & Y0         & 350 (350$-$400)           & 4.75 (4.75$-$5.00)  & 4       & 0.93 (0.86$-$0.94) & 20 (20$-$30)  \\ 
\Weighteentwentyeight\    & $>$Y0     & $\leq$300                 & $\cdots$            & $\cdots$ & $\cdots$           & $\cdots$            \\ 
\Wtwentyfiftysix\         & Y0         & 350 (350$-$400)           & 4.75 (4.50$-$5.00)  & 4       & 0.93 (0.86$-$1.01) & 20 (12$-$30)  \\ 

\enddata

\tablecomments{The parameters for the best fitting Marley \& Saumon
  models are listed and the range of parameters consistent with the data
  is given in parentheses.  The effective temperature limit for
  \Weighteentwentyeight\ was estimated by identifying those models with
  peak $J$-band fluxes equal to or less than the peak flux in the $H$
  band and by comparing the observed $J-$W2 color to model $J-$W2
  colors.}

\end{deluxetable}

%
%=========================================  Distance Estimates ========================================================
%
\begin{deluxetable}{rlccc}
\tablecolumns{6}
\tabletypesize{\scriptsize} 
\tablewidth{0pc}
\tablecaption{\label{tab:dist}Distance Estimates}
\tablehead{
\colhead{Object} & 
\colhead{SpType} & 
\colhead{d$_\mathrm{spec}$ (pc)\tablenotemark{a}} & 
\colhead{d$_\mathrm{\pi}$ (pc)\tablenotemark{b}} & 
\colhead{d$_\mathrm{phot}$ (pc)}\tablenotemark{c}}

\startdata
\UGPSzeroseventwotwo\     & T9         & 11.1 (10.4$-$11.1) & 3.6$-$4.7 &  $\cdots$ \\
\Wzeroonefoureight\       & T9.5       & 14.7 (13.1$-$14.7) & $\cdots$  &  12.1 \\
\Wzerofourten\            & Y0         & 7.1 (3.3$-$8.7)    & $\cdots$  &  9.0 \\  
\Wfourteenzerofive\       & Y0 (pec?)  & 3.8                & $\cdots$  &  8.6 \\  
\Wfifteenfortyone\        & Y0         & 1.8 (1.8$-$2.0)    & 2.2$-$4.1 &  8.2 \\
\Wseventeenthirtyeight\   & Y0         & 3.4 (3.4$-$7.3)    & $\cdots$  &  10.5 \\
\Weighteentwentyeight\    & $>$Y0      & $\cdots$           & $\cdots$  &  $<$9.4 \\
\Wtwentyfiftysix\         & Y0         & 3.0 (2.4$-$6.4)    & $\cdots$  &  7.7 \\

\enddata

\tablenotetext{a}{Spectroscopic distance estimates derived as described
  in \S\ref{sec:atmosprop}.  The distance corresponding to the best
  fitting model is given and the range of distances corresponding to
  models that are consistent with the data are given in parentheses.}

\tablenotetext{b}{Parallactic distance for \UGPSzeroseventwotwo\ and
  \Wfifteenfortyone\ from \citet{2010MNRAS.408L..56L} and
  \citet{Davy11}, respectively.}

\tablenotetext{c}{Photometric distance estimates from \citet{Davy11}.}

\end{deluxetable}

\end{document}